\documentclass[reprint,amsmath,amssymb,aps,prb]{revtex4-2}
\PassOptionsToPackage{obeyFinal}{todonotes}
\usepackage[latin9]{inputenc}
\usepackage{todonotes}
\usepackage{graphicx}
\usepackage{ulem}
\usepackage{booktabs}

\makeatletter
\usepackage{dcolumn}
\usepackage{bm}
\usepackage{xcolor}

\usepackage{etoolbox}

\makeatother

\begin{document}
\title{Possible Spatial Correlation between Superconducting and Pseudogap Dynamics in a Bi-based Cuprate}
\author{T. Shimizu}
\affiliation{Department of Applied Physics, Hokkaido University, Sapporo 060-8628,
Japan.}
\author{T. Kurosawa}
\affiliation{Department of Applied Physics, Hokkaido University, Sapporo 060-8628,
Japan.}
\author{S. Tsuchiya}
\affiliation{Department of Applied Physics, Hokkaido University, Sapporo 060-8628,
Japan.}
\author{R. Tobise}
\affiliation{Department of Applied Physics, Hokkaido University, Sapporo 060-8628,
Japan.}
\author{K. Yamane}
\affiliation{Department of Applied Physics, Hokkaido University, Sapporo 060-8628,
Japan.}
\author{R. Morita}
\affiliation{Department of Applied Physics, Hokkaido University, Sapporo 060-8628,
Japan.}
\author{Y. Toda}
\affiliation{Department of Applied Physics, Hokkaido University, Sapporo 060-8628,
Japan.}
\author{M. Oda}
\affiliation{Department of Physics, Hokkaido University, Sapporo 060-0810, Japan.}

\date{\today}
\begin{abstract}
Understanding the interplay between superconductivity and the pseudogap phase is essential for elucidating the mechanism of high-temperature superconductivity in cuprates.
Here we provide direct spatial evidence that these two states are locally and intrinsically correlated.
Using spatially and temporally resolved measurements of photoinduced quasiparticle dynamics in optimally doped Bi$_2$Sr$_{1.7}$La$_{0.3}$CuO$_{6+\delta}$ (La-Bi2201), we reveal micrometer-scale spatial contrasts in the transient reflectivity that arise from local variations in the threshold fluence required to disrupt either the superconducting or pseudogap state.
The superconducting response remains spatially uniform, whereas the pseudogap exhibits intrinsic inhomogeneity, yet the spatial variations of their threshold fluences closely track each other, establishing a robust local correlation between the two.
These results introduce a bulk-sensitive ultrafast optical methodology for visualizing hidden spatial correlations in correlated materials and provide new benchmarks for understanding the intertwined phases in cuprates.
\end{abstract}
\maketitle

\section{Introduction}

High-temperature superconductivity (SC) in cuprates remains a central challenge in condensed matter physics. A hallmark of these materials is the emergence of the pseudogap (PG) state, which develops above the superconducting transition temperature $T_{\rm c}$ and coexists with superconductivity below $T_{\rm c}$~\cite{timusk1999,norman1998,keimer2015,hashimoto2014}. Understanding the interplay between these two states has been debated for decades.

Momentum-resolved probes such as angle-resolved photoemission spectroscopy (ARPES) have revealed that the PG opens in the antinodal region with $d$-wave symmetry, while the superconducting gap develops near the nodes below $T_{\rm c}$~\cite{tanaka2006,lee2006,kondo2007,okada2008}. This dichotomy implies depletion of carrier density by the PG and competition with SC. Nanoscale imaging with scanning tunneling microscopy/spectroscopy (STM/STS) further shows that regions with pronounced PG features often exhibit suppressed superconducting coherence peaks, reinforcing this competitive picture~\cite{gomes2007,Vershinin2004,kohsaka2008}. Conversely, other systematic studies, including optical conductivity and ARPES on optimally doped cuprates, have suggested cooperative scaling relations between the PG and $T_{\rm c}$~\cite{ideta2010,tajima2024}. More recently, tunneling spectroscopy has indicated that the PG energy scale evolves in concert with superconducting pairing correlations~\cite{niu2024,ye2023}. These contrasting results underscore the need for experimental approaches that can capture both PG and SC from complementary perspectives.

Time-resolved optical spectroscopy offers such a perspective by directly probing nonequilibrium quasiparticle (QP) dynamics in both phases~\cite{de2021,demsar1999,kaindl2000,luo2006}. Unlike ARPES and STM, this all-optical pump-probe method is inherently bulk-sensitive and can track the temporal evolution of photoinduced dynamics to reveal how the PG and SC states interact during formation and recovery~\cite{toda2011,coslovich2013, toda2023,akiba2024}. Importantly, the method requires no lithography or electrode patterning and can be extended to spatial mapping of transient optical responses. Bulk-sensitive, spatially resolved ultrafast spectroscopy thus provides a powerful and versatile approach for detecting intrinsic inhomogeneities and hidden spatial correlations in correlated electron systems.

To investigate the relationship between the PG and SC states under conditions where $T_{\rm c}$ is maximized, we employ a combined spatially and temporally resolved pump-probe reflectivity method focusing on an optimally doped single-layer Bi$_2$Sr$_{1.7}$La$_{0.3}$CuO$_{6+\delta}$ (La-Bi2201).
Because La-Bi2201 has a relatively low maximum $T_{\rm c}$, the quasiparticle relaxation dynamics associated with the SC and PG responses are clearly separated in time, and their temperature- and fluence-dependent behaviors are well defined, enabling selective separation and analysis of the two responses.

By performing one-dimensional (1D) line scans and two-dimensional (2D) imaging of photoinduced reflectivity changes, we reveal micrometer-scale spatial variations in the transient signals. Fluence-dependent measurements at representative locations identify local variations in the threshold fluence required to disrupt either the SC or PG state.
Furthermore, we find that these thresholds show systematic correspondence with  $T_{\rm c}$ and the PG energy, respectively,
and that the spatial variation of the SC threshold fluence closely parallels that of the PG threshold fluence. Our results highlight the capability of spatially resolved ultrafast optical spectroscopy to disentangle and correlate competing electronic orders, providing a new approach to explore the interplay between superconductivity and the pseudogap in cuprates.

\section{Experimental}

Single crystals of rare-earth substituted R-Bi2201 (R = La and Eu) with R content of 0.3 were grown under 1~atm flowing oxygen, corresponding to the optimally doped regime reported previously~\cite{eisaki2004,kurosawa2010,kurosawa2016}. 
The superconducting transition temperatures determined from magnetic susceptibility measurements were $T_{\rm c} \approx 34$ K for La-Bi2201 and $T_{\rm c} \approx 20$ K for Eu-Bi2201.

Time-resolved pump-probe spectroscopy was performed using a cavity-dumped Ti:Al$_2$O$_3$ laser oscillator (pulse duration 120~fs and $\lambda_{\rm pr}=800$~nm) for the probe and its second harmonic ($\lambda_{\rm P}=400$~nm) for the pump pulses. The repetition rate was set to 270~kHz to suppress steady-state heating. 
Magnetic susceptibility measurements confirmed that any temperature increase during the measurements was less than 1~K.

The sample was mounted in a helium-flow cryostat and precisely positioned using a motorized $XY$ translation stage. All optical pulses were linearly polarized, coaxially combined, and focused onto the sample through an objective lens. Spatially resolved measurements were achieved with a spatial resolution of approximately 5~$\mu$m. The transient reflectivity signal $\Delta R/R$, representing the relative change in reflectivity of the probe, was measured using a photodetector and a lock-in amplifier synchronized with a mechanically modulated pump beam.

\section{Results}

\begin{figure}[htbp]
\includegraphics[width=0.98\columnwidth]{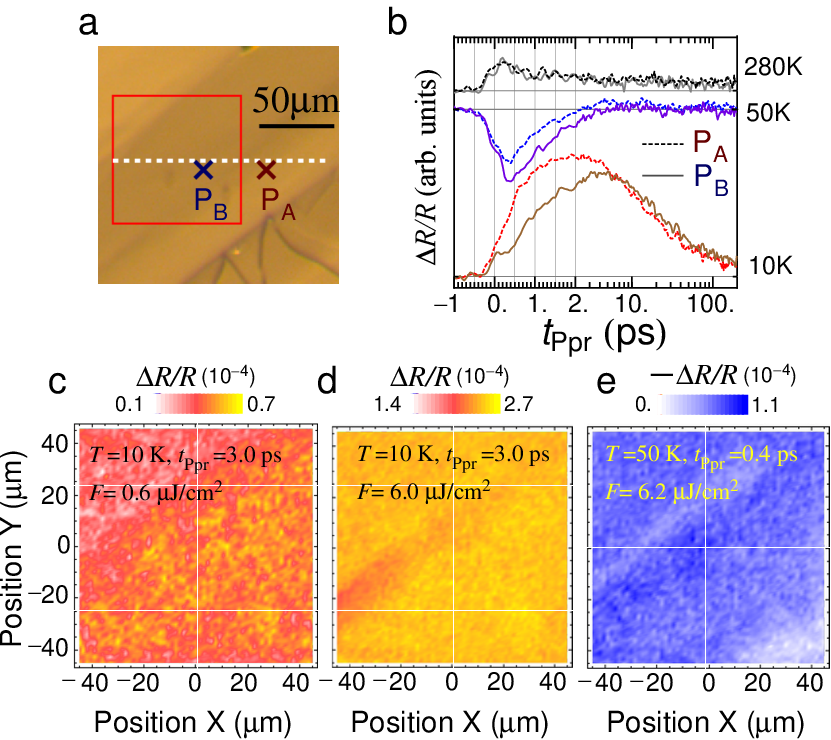}
\caption{(Color online)
(a) Optical microscope image of the sample surface. The red box marks the two-dimensional (2D) scan region, while the white dashed line indicates the one-dimensional (1D) scan path.
(b) Representative transient reflectivity changes $\Delta R/R$, recorded at selected temperatures and at two distinct positions, P$_{\rm A}$ (dashed) and P$_{\rm B}$ (solid), as indicated by the crosses in (a), under a pump fluence of ${\mathcal{F}}=15~\mu$J/cm$^{2}$. For clarity, traces at different temperatures are vertically offset.
(c),(d) 2D images of $\Delta R/R$ over a $90 \times 90~\mu\mathrm{m}^2$ area corresponding to the region indicated by the red box in (a), acquired at $T=10~\mathrm{K}$ and a probe delay of $t_{\rm Ppr}=3.0$~ps with pump fluences of ${\mathcal{F}}=0.6$ and $6.0~\mu$J/cm$^{2}$, respectively. 
At this probe delay the PG contribution has largely relaxed (see panel (b)), so that the signal primarily reflects the superconducting component $\Delta R_{\mathrm{SC}}/R$.
(e) 2D map of the PG response measured at $T=50~\mathrm{K}$, $t_{\rm Ppr}=0.4$~ps, and ${\mathcal{F}}=6.2~\mu$J/cm$^{2}$. Because $\Delta R_{\mathrm{PG}}/R$ is negative, $-\Delta R/R$ is plotted.
}
\label{fig_2D}
\end{figure}

We first outline the transient reflectivity change, $\Delta R/R$, observed in La-Bi2201. Figure~\ref{fig_2D}(b) shows representative $\Delta R/R$ transients measured at two distinct positions, P$_{\rm A}$ and P$_{\rm B}$, under a pump fluence of $\mathcal{F}=15~\mu$J/cm$^2$. The corresponding positions on the microscope image are marked by crosses in Fig.~\ref{fig_2D}(a). The difference between the signals at these two positions will be discussed later. For clarity, the transients are vertically offset according to the three selected temperatures.
Below $T_{\mathrm{c}}$, the $\Delta R/R$ response is dominated by QP relaxation dynamics associated with superconductivity ($\Delta R_{\mathrm{SC}}/R$). The relaxation of SC QPs on the timescale of several tens of picoseconds reflects their recombination across the SC gap.
Above $T_{\mathrm{c}}$, the dynamics are governed by PG QPs ($\Delta R_{\mathrm{PG}}/R$), which exhibit an opposite sign compared with the SC response. The relaxation of PG QPs occurs on a timescale of $\lesssim$1~ps, faster than that of SC QPs and associated with the partial-gap nature of the PG.
Well above $T_{\mathrm{c}}$ (and above the PG onset), metallic electron-phonon relaxation dominates $\Delta R/R$ ($\Delta R_{\mathrm{ER}}/R$).
These three characteristic transient components have been widely reported in cuprate superconductors~\cite{demsar1999,Dvorsek2002,Segre2002,Kusar2005,toda2011,toda2021,akiba2024}.

Figures~\ref{fig_2D}(c)-\ref{fig_2D}(e) present 2D spatial distributions of $\Delta R/R$ over an area of $90 \times 90~\mu\mathrm{m}^2$, corresponding to the region in Fig.~\ref{fig_2D}(a). In Figs.~\ref{fig_2D}(c) and \ref{fig_2D}(d), recorded at $T=10~\mathrm{K}$ and $t_{\rm Ppr}=3.0$~ps with ${\mathcal{F}}=0.6$ and $6.0~\mu$J/cm$^{2}$, respectively, the signal primarily reflects $\Delta R_{\mathrm{SC}}/R$. 
At this probe delay, the PG contribution has largely relaxed, as seen in the transient traces in Fig.~\ref{fig_2D}(b), so that the spatial variation mainly reflects the superconducting component.
In contrast, Fig.~\ref{fig_2D}(e) shows the distribution obtained at $T=50~\mathrm{K}$ and $t_{\rm Ppr}=0.4$~ps with ${\mathcal{F}}=6.2~\mu$J/cm$^{2}$, where $\Delta R/R$ is dominated by $\Delta R_{\mathrm{PG}}/R$ and is plotted as $-\Delta R/R$ for clarity. Notably, with increasing pump fluence, the initially gradual spatial variation of $\Delta R_{\mathrm{SC}}/R$ in Fig.~\ref{fig_2D}(c) evolves into a more well-defined profile in Fig.~\ref{fig_2D}(d). This profile qualitatively resembles the characteristic pattern of $\Delta R_{\mathrm{PG}}/R$ in Fig.~\ref{fig_2D}(e), although their variations exhibit an anticorrelated.

\begin{figure}[htbp]
\includegraphics[width=0.98\columnwidth]{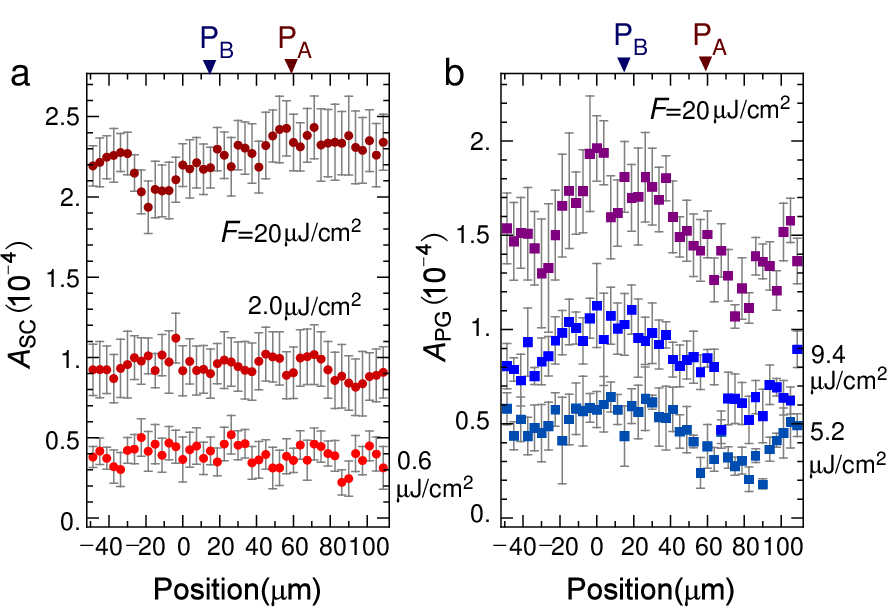}
\caption{(Color online)
1D spatial distributions of $A_{\rm SC}$ and $A_{\rm PG}$ at different pump fluences, extracted from $\Delta R/R$ transients measured at 43 positions along the dashed line in Fig.~\ref{fig_2D}(a) with spacing $\Delta x = 3.75~\mu$m. $A_{\rm SC}$ is defined as $\langle \Delta R/R \rangle_{2-10~\mathrm{ps}}$ at $T=10~\mathrm{K}$, whereas $A_{\rm PG}$ is obtained from $\langle -\Delta R/R \rangle_{0.1-0.5~\mathrm{ps}}$ at $T=50~\mathrm{K}$.
}
\label{fig_1D}
\end{figure}

The fluence-dependent relationship between $\Delta R_{\mathrm{SC}}/R$ and $\Delta R_{\mathrm{PG}}/R$ is more clearly revealed in Fig.~\ref{fig_1D}, which shows spatial distributions of the SC and PG components at different pump fluences (the fluences in Fig.~\ref{fig_2D} were chosen for visual clarity, whereas Fig.~\ref{fig_1D} presents the systematic set used for quantitative analysis). These distributions were extracted from $\Delta R/R$ transients measured along the $\sim$160~$\mu$m dashed line in Fig.~\ref{fig_2D}(a). Further details are provided in the Appendix A. To quantify $\Delta R_{\mathrm{SC,PG}}/R$, we plot the SC response amplitude $A_{\rm SC}\equiv\langle \Delta R/R \rangle_{2-10~\mathrm{ps}}$ at $T=10~\mathrm{K}$ and the PG response amplitude $A_{\rm PG}\equiv\langle -\Delta R/R \rangle_{0.1-0.5~\mathrm{ps}}$ at $T=50~\mathrm{K}$.
At weak excitation [Fig.~\ref{fig_1D}(a)], $A_{\rm SC}$ is spatially uniform within experimental uncertainty. The apparent modulation in Fig.~\ref{fig_2D}(c) is enhanced by the color scale, whereas the quantitative analysis in Fig.~\ref{fig_1D}(a) confirms near uniformity. With increasing fluence, micron-scale modulation becomes evident. In contrast, $A_{\rm PG}$ exhibits spatial modulation already at weak excitation [Fig.~\ref{fig_1D}(b)] and is approximately complementary to $A_{\rm SC}$ at high-fluence excitation.

Below $T_{\rm c}$, under strong excitation where the superconducting condensate is significantly suppressed, the transient reflectivity contains contributions from both the SC and PG components. 
In such a situation, it is not straightforward to disentangle the two components solely from the spatial profiles of $A_{\rm SC}$ and $A_{\rm PG}$. 
On the other hand, their fluence dependences reflect the intrinsic characteristics of each component.
In particular, the deviation from the linear response in the low-fluence regime defines characteristic threshold fluences associated with the suppression of the SC condensate and the collapse of the PG response. 
As shown below and further supported by Appendix D, these threshold fluences provide a useful metric for discussing the relationship between the SC and PG components in the spatial analysis.

\begin{figure}[htbp]
\includegraphics[width=0.98\columnwidth]{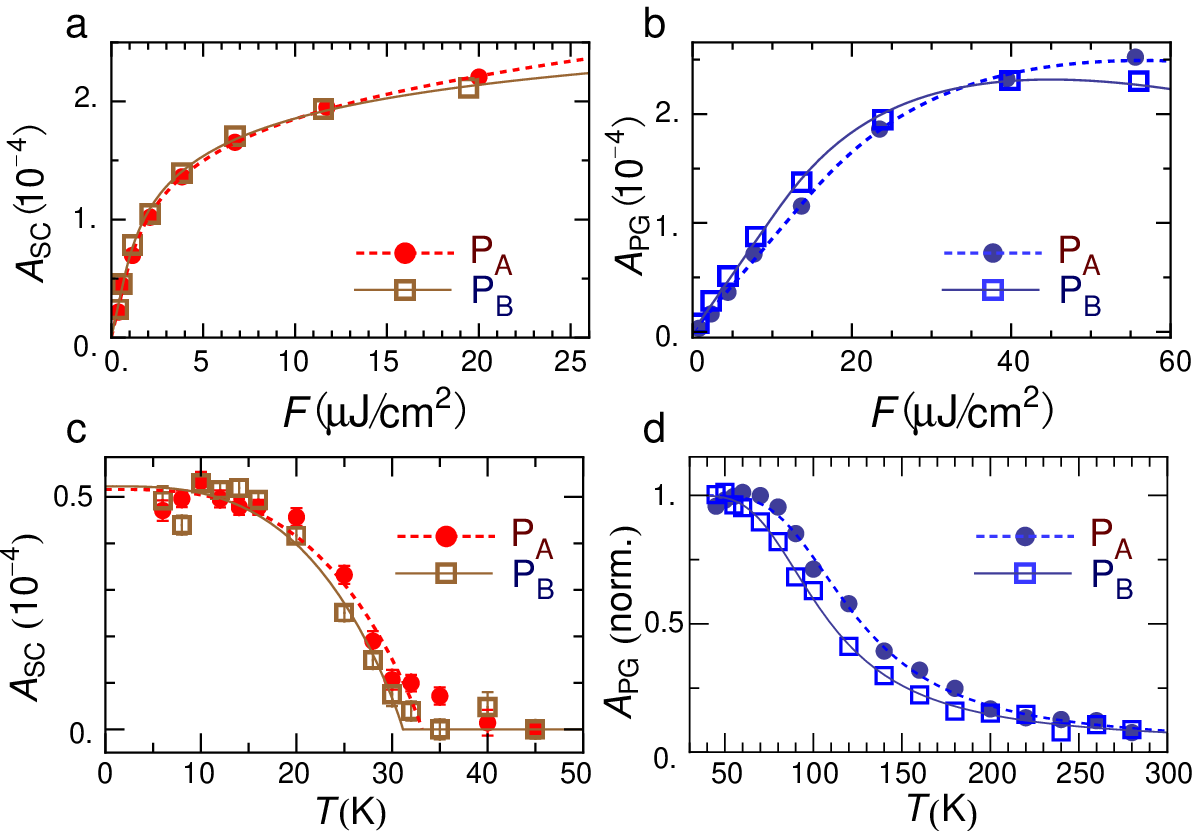}
\caption{(Color online)
Fluence dependence of SC and PG response amplitudes for (a) $A_{\rm SC}$ and (b) $A_{\rm PG}$ at positions P$_{\rm A}$ and P$_{\rm B}$ in Fig.~\ref{fig_2D}(a). Dashed (P$_{\rm A}$) and solid (P$_{\rm B}$) lines represent fits based on the finite-penetration-depth excitation model~\cite{kusar2008,naseska2018}, from which the threshold fluences $\mathcal{F}_{\rm th}^{\rm SC}$ and $\mathcal{F}_{\rm th}^{\rm PG}$ are extracted.
Temperature dependence of (c) $A_{\rm SC}$ and (d) $A_{\rm PG}$. Dashed (P$_{\rm A}$) and solid (P$_{\rm B}$) lines correspond to fits using (c) the Mattis-Bardeen formula with a BCS-like gap~\cite{mertelj2009distinct} and (d) a temperature-independent gap model~\cite{kabanov1999}, respectively.
}
\label{fig_AB}
\end{figure}

\begin{table*}[t]
\centering
\caption{(Color online) Threshold fluences for superconducting ($\mathcal{F}_{\mathrm{th}}^{\mathrm{SC}}$)
and pseudogap ($\mathcal{F}_{\mathrm{th}}^{\mathrm{PG}}$) responses, superconducting transition temperature ($T_{\mathrm{c}}$), and pseudogap energy ($\Delta_{\mathrm{PG}}$) at positions P$_{\rm A}$ and P$_{\rm B}$.}
\label{tab:thresholds}
\begin{tabular}{ccccc}
\toprule
Position & $\mathcal{F}_{\mathrm{th}}^{\mathrm{SC}}$ ($\mu$J/cm$^2$) & $T_{\mathrm{c}}$ ($\mathrm{K}$) & $\mathcal{F}_{\mathrm{th}}^{\mathrm{PG}}$ ($\mu$J/cm$^2$) & $\Delta_{\mathrm{PG}}$ (meV) \\
\midrule
P$_{\rm A}$ & $0.63 \pm 0.06$ & $33.3 \pm 0.6$ & $13.1 \pm 0.6$ & $45.8 \pm 4.2$ \\
P$_{\rm B}$ & $0.59 \pm 0.06$ & $31.3 \pm 0.6$ & $9.6 \pm 0.5$ & $37.1 \pm 2.6$ \\
\bottomrule
\end{tabular}
\end{table*}

Figures~\ref{fig_AB}(a) and \ref{fig_AB}(b) show the fluence dependence of $A_{\rm SC}$ and $A_{\rm PG}$, respectively, for P$_{\rm A}$ and P$_{\rm B}$. In general, the $\Delta R/R$ response in the SC (PG) state can be divided into a linear regime, where $\Delta R/R$ increases proportionally with the density of photoinduced QPs, and a nonlinear (saturated) regime, where further fluence increase does not yield a proportional response because the SC (PG) phase is partially destroyed within the excited volume. The fluence at which the response departs from linearity defines the phase-destruction threshold $\mathcal{F}_{\rm th}^{\rm SC,PG}$, which reflects the energy required to destroy the corresponding phase. As shown by the dashed (P$_{\rm A}$) and solid (P$_{\rm B}$) lines, the finite-penetration-depth excitation model~\cite{kusar2008,naseska2018} reproduces the observed fluence dependence well, with $\mathcal{F}_{\rm th}^{\rm SC,PG}$ serving as an effective fitting parameter (details in the Appendix B and D).

In Fig.~\ref{fig_AB}(a), $A_{\rm SC}(\mathcal{F})$ at P$_{\rm A}$ and P$_{\rm B}$ differs only slightly; their $\mathcal{F}_{\rm th}^{\rm SC}$ values differ by $\sim$7\% (Table~\ref{tab:thresholds}). This small difference is also discernible in the $\Delta R/R$ traces at $T=10~\mathrm{K}$ under excitation above the superconducting threshold, $\mathcal{F}=15~\mu$J/cm$^2>\mathcal{F}_{\rm th}^{\rm SC}$ [Fig.~\ref{fig_2D}(b)]: the rise of $\Delta R/R$ at P$_{\rm B}$ is suppressed compared with P$_{\rm A}$, indicating a stronger early-time PG contribution at P$_{\rm B}$.

For complementary analysis, the temperature dependence of $A_{\rm SC}$ and $A_{\rm PG}$ is shown in Figs.~\ref{fig_AB}(c) and \ref{fig_AB}(d), respectively. The lines in Fig.~\ref{fig_AB}(c) are fits to the Mattis-Bardeen model with $T_{\rm c}$ as a parameter~\cite{mertelj2009distinct}, while those in Fig.~\ref{fig_AB}(d) correspond to fits to a temperature-independent gap model with $\Delta_{\rm PG}$ as a parameter~\cite{kabanov1999}. The corresponding $\Delta R/R$ data are provided in Appendix~C.

The results of these analyses are summarized in Table~\ref{tab:thresholds}. Previous studies have shown that $\mathcal{F}_{\rm th}^{\rm SC}$ scales with $T_{\rm c}^{\,2}$ across various high-$T_{\rm c}$ superconductors~\cite{stojchevska2011}, while $\mathcal{F}_{\rm th}^{\rm PG}$ in Bi2212 scales with doping level~\cite{madan2015}, suggesting a correlation with $\Delta_{\rm PG}$.
Table~\ref{tab:thresholds} reflects these trends: the relative magnitudes of $T_{\rm c}$ for P$_{\rm A}$ and P$_{\rm B}$ are consistent with those of $\mathcal{F}_{\rm th}^{\rm SC}$, and the magnitudes of $\Delta_{\rm PG}$ correspond to $\mathcal{F}_{\rm th}^{\rm PG}$. 
Furthermore, comparison between P$_{\rm A}$ and P$_{\rm B}$ indicates a correspondence between $\mathcal{F}_{\rm th}^{\rm SC}$ and $\mathcal{F}_{\rm th}^{\rm PG}$.

\begin{figure}[htbp]
\includegraphics[width=0.98\columnwidth]{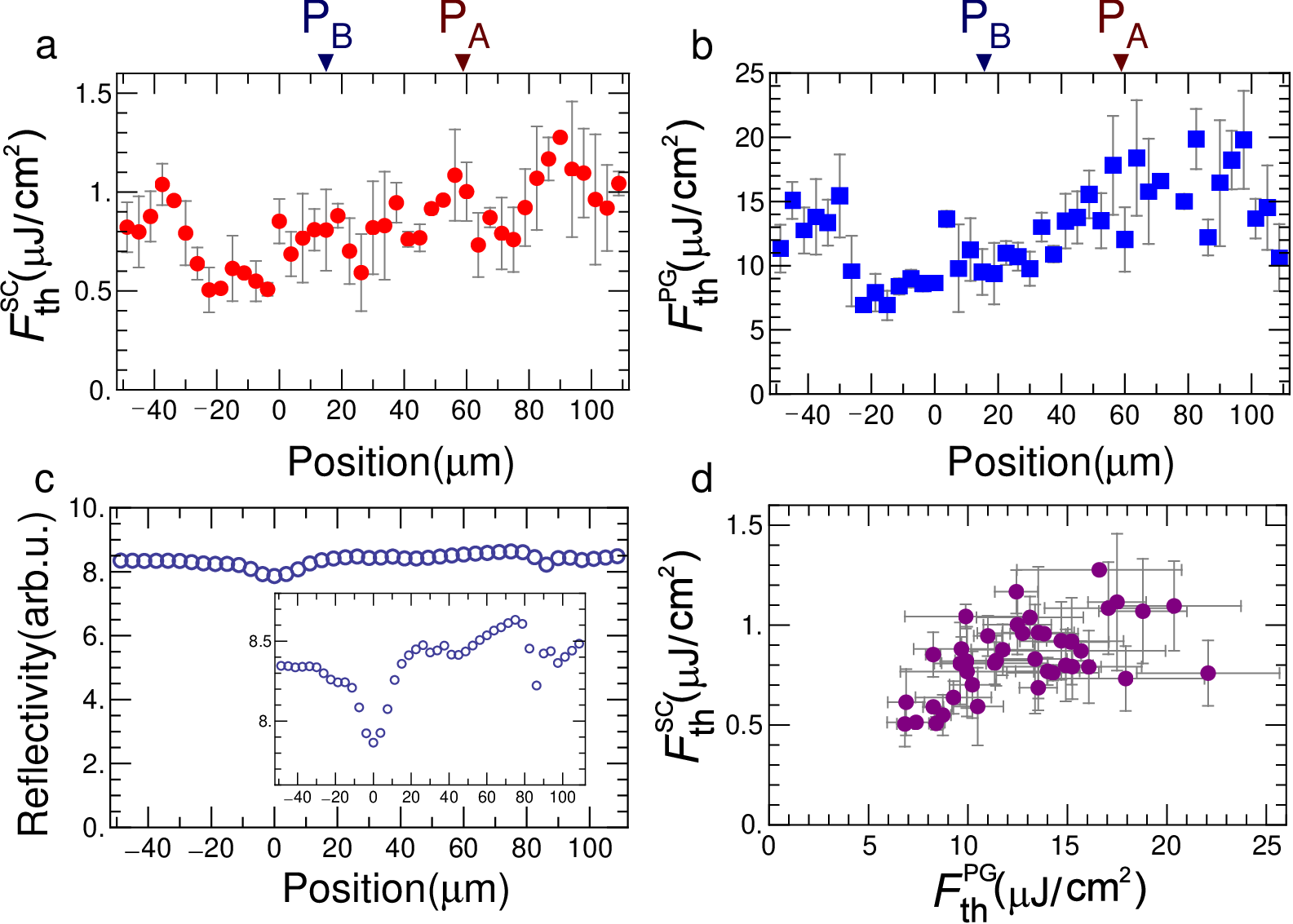}
\caption{(Color online)
Spatial distributions of phase-destruction thresholds for (a) the superconducting phase, ${\mathcal{F}}_{\rm th}^{\rm SC}$ at $T=10~\mathrm{K}$, and (b) the PG phase, ${\mathcal{F}}_{\rm th}^{\rm PG}$ at $T=50~\mathrm{K}$. Thresholds are obtained from the fluence dependence of $\Delta R/R$ measured along the dashed line in Fig.~\ref{fig_2D}(a). (c) Corresponding spatial distribution of the reflectivity (inset: magnified view). (d) Correlation between ${\mathcal{F}}_{\rm th}^{\rm SC}$ and ${\mathcal{F}}_{\rm th}^{\rm PG}$ from (a) and (b).
}
\label{fig_Fth1D}
\end{figure}

We extend this comparison to multiple positions along the 1D path. Figures~\ref{fig_Fth1D}(a) and \ref{fig_Fth1D}(b) show the corresponding spatial distributions of $\mathcal{F}_{\rm th}^{\rm SC}$ and $\mathcal{F}_{\rm th}^{\rm PG}$, extracted from fluence dependences of $A_{\rm SC}$ at $T=10~\mathrm{K}$ and $A_{\rm PG}$ at $T=50~\mathrm{K}$ using the same time-domain analysis (the corresponding data are provided in Appendix~B). The error bars represent standard errors from the nonlinear fits. For reference, the spatial variation in steady-state reflectivity is shown in Fig.~\ref{fig_Fth1D}(c) (inset: magnified view). Figures~\ref{fig_Fth1D}(a) and \ref{fig_Fth1D}(b) exhibit distinct spatial features uncorrelated with reflectivity (absorption) but closely resembling each other, consistent with the spatial distributions in Figs.~\ref{fig_1D} and \ref{fig_2D}(d),(e). For example, a sharp transition near $X\sim-20~\mu$m is commonly observed. Although the reflectivity in Fig.~\ref{fig_Fth1D}(c) shows a weak overall trend similar to Figs.~\ref{fig_Fth1D}(a),(b), its amplitude is much smaller and likely reflects minor surface or optical inhomogeneity. The correlation between Figs.~\ref{fig_Fth1D}(a) and \ref{fig_Fth1D}(b) is further supported by Fig.~\ref{fig_Fth1D}(d), where data from all positions reveal a nearly linear relationship between $\mathcal{F}_{\rm th}^{\rm SC}$ and $\mathcal{F}_{\rm th}^{\rm PG}$.

\begin{figure}[htbp]
\includegraphics[width=0.98\columnwidth]{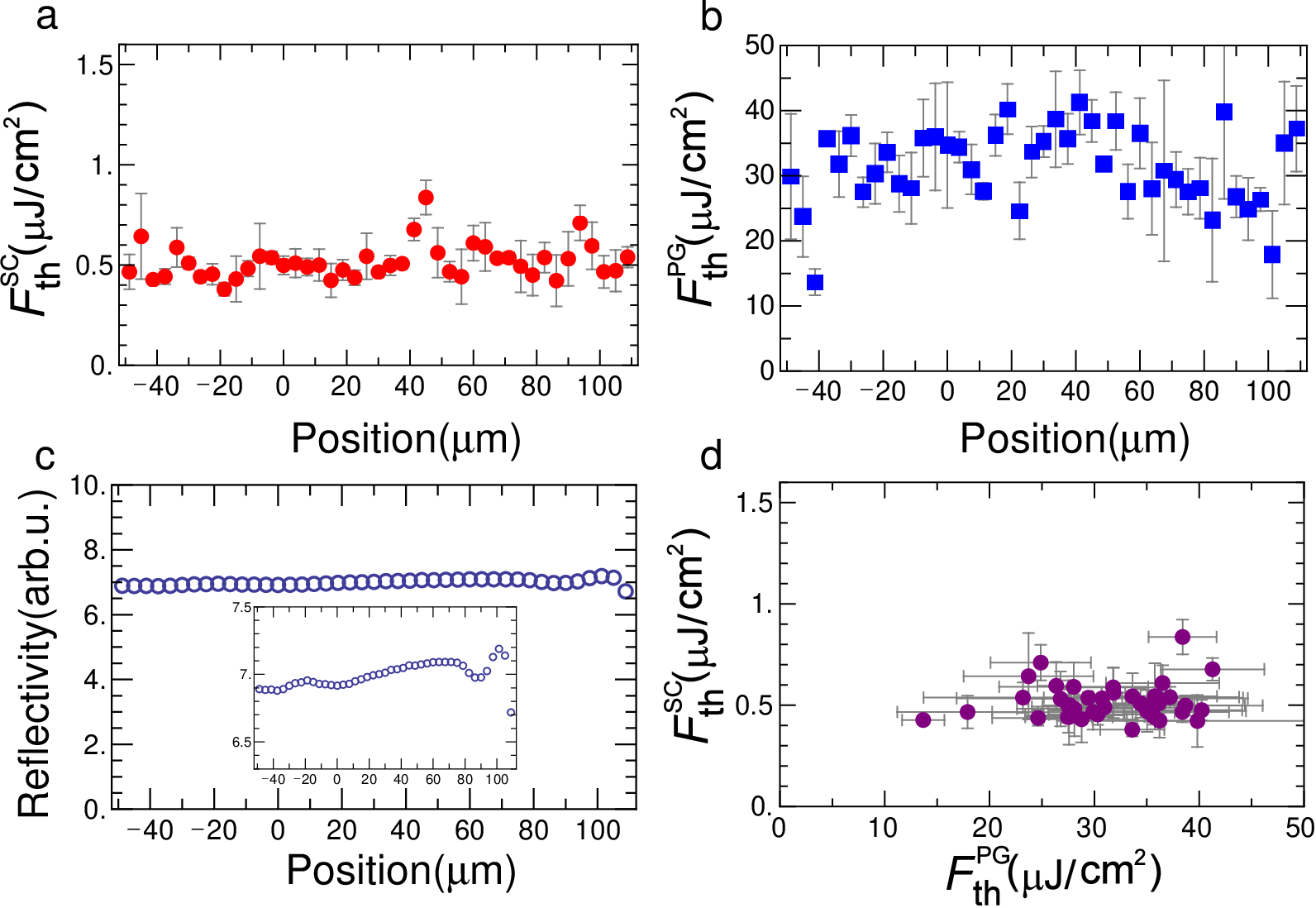}
\caption{(Color online)
Spatial distributions of the phase-destruction thresholds in Eu-Bi2201. (a) ${\mathcal{F}}_{\rm th}^{\rm SC}$ at $T=10~\mathrm{K}$, (b) ${\mathcal{F}}_{\rm th}^{\rm PG}$ at $T=50~\mathrm{K}$, and (c) reflectivity (inset: magnified view). (d) Correlation between ${\mathcal{F}}_{\rm th}^{\rm SC}$ and ${\mathcal{F}}_{\rm th}^{\rm PG}$ from (a) and (b).
}
\label{fig_Fth1DEu}
\end{figure}

For comparison, we performed the same analysis on a Eu-substituted Bi2201 sample (Eu-Bi2201, $T_{\rm c}\approx20$ K).
In this sample, $\mathcal{F}_{\rm th}^{\rm SC}$ (Fig.~\ref{fig_Fth1DEu}(a)) decreases compared with La-Bi2201, whereas $\mathcal{F}_{\rm th}^{\rm PG}$ (Fig.~\ref{fig_Fth1DEu}(b)) increases. This behavior is broadly consistent with the fact that Eu substitution is known to enhance disorder in the crystal and suppress the maximum $T_{\rm c}$.
For reference, the spatial variation in steady-state reflectivity is shown in Fig.~\ref{fig_Fth1DEu}(c).
Furthermore, the spatial variation of $\mathcal{F}_{\rm th}^{\rm SC}$ is significantly reduced compared with that in La-Bi2201, and the correlation between $\mathcal{F}_{\rm th}^{\rm SC}$ and $\mathcal{F}_{\rm th}^{\rm PG}$ that was clearly observed in La-Bi2201 becomes less pronounced (Fig.~\ref{fig_Fth1DEu}(d)).
These results suggest that the correlation between the SC and PG thresholds depends on the sample conditions, and is reduced in the Eu-substituted system.

\section{Discussion}
Below $T_{\rm c}$, strong excitation suppresses the superconducting condensate, and the transient reflectivity signal contains contributions from both the superconducting (SC) and pseudogap (PG) responses. This makes it difficult to separate the two components based solely on the spatial distribution at a fixed delay time.
To address this issue, we carried out a 1D spatially resolved analysis based on the time evolution and fluence dependence of the transient response. This approach enables a reliable determination of the threshold fluence for the destruction of the superconducting condensate, $\mathcal{F}_{\rm th}^{\rm SC}$, even under conditions where SC and PG responses coexist below $T_{\rm c}$.
Importantly, in the La-Bi2201 sample, the threshold fluences $\mathcal{F}_{\rm th}^{\rm SC}$ and $\mathcal{F}_{\rm th}^{\rm PG}$ exhibit a strong spatial correlation, whereas this correlation is nearly absent in the Eu-Bi2201 sample. These observations suggest that the correlation between the SC and PG thresholds becomes most evident under conditions where $T_{\rm c}$ is maximized.

This study reveals a spatial correlation between $\mathcal{F}_{\rm th}^{\rm SC}$ and $\mathcal{F}_{\rm th}^{\rm PG}$ in La-Bi2201. As shown in Figs.~\ref{fig_Fth1D}(a),(b), both thresholds exhibit closely matched spatial variations across $\sim$160~$\mu$m, while the simultaneously measured reflectivity remains nearly uniform and uncorrelated [Fig.~\ref{fig_Fth1D}(c)]. The point-by-point plot in Fig.~\ref{fig_Fth1D}(d) further reveals a near-linear positive relationship, indicating that regions with higher $\mathcal{F}_{\rm th}^{\rm SC}$ also possess higher $\mathcal{F}_{\rm th}^{\rm PG}$.

The spatial distributions of the threshold fluences $\mathcal{F}_{\rm th}^{\rm SC,PG}$ and amplitudes $A_{\rm SC,PG}$ in La-Bi2201 provide insight into distinct characteristics of SC and PG. In Fig.~\ref{fig_1D}(a), $A_{\rm SC}$ remains spatially uniform under weak excitation but develops pronounced spatial variations under high-fluence excitation, correlating with $\mathcal{F}_{\rm th}^{\rm SC}$ in Fig.~\ref{fig_Fth1D}(a). This indicates that the QP density associated with SC is homogeneous on the micrometer scale, whereas the variations in $\mathcal{F}_{\rm th}^{\rm SC}$ and $A_{\rm SC}$ under high-fluence excitation reflect local variations in condensate stability. By contrast, $A_{\rm PG}$ in Fig.~\ref{fig_1D}(b) exhibits clear inhomogeneity even in the weak-excitation regime and anticorrelates with $\mathcal{F}_{\rm th}^{\rm PG}$ [Fig.~\ref{fig_Fth1D}(b)], implying that larger $A_{\rm PG}$ corresponds to a lower PG energy scale~\cite{Vershinin2004,kurosawa2016}.

The spatial inhomogeneity observed here in La-Bi2201 does not directly reflect the nanoscale pseudogap inhomogeneity reported by STM. Due to the micrometer-scale spatial resolution of the present measurements, we instead probe long-wavelength spatial modulations that arise from spatial averaging over many nanoscale regions. Such background variations may be influenced by gradual spatial modulations of local doping, residual strain, or the distribution of defects in the crystal, as discussed further below.

The 2D images in Figs.~\ref{fig_2D}(d),(e) reveal micrometer-scale modulations along a crystallographic axis, demonstrating that the QP response is not uniform. The microscopic origin of these modulations cannot be identified directly from our measurements. One possible source is an inhomogeneous distribution of doping. Since $\Delta_{\rm PG}$ decreases monotonically with increasing doping, the observed proportionality between $\mathcal{F}_{\rm th}^{\rm SC}$ and $\mathcal{F}_{\rm th}^{\rm PG}$ can be consistently explained if the optimally doped sample is situated on the overdoped side of the phase diagram~\cite{kondo2009,kurosawa2010,kurosawa2016}.

Another possible origin is structural disorder. The sample is tuned to optimal doping by out-of-plane disorder introduced through La-Sr substitution~\cite{kurosawa2016}. Previous STM and ARPES studies reported an anticorrelation between PG and SC energy scales across different substitutional series, attributed to variations in the ionic radius of the out-of-plane element~\cite{kurosawa2016,okada2008,okada2011}. In contrast, our data exhibit a positive correlation between $\mathcal{F}_{\rm th}^{\rm PG}$ and $\mathcal{F}_{\rm th}^{\rm SC}$, suggesting a different mechanism: because the present measurements use a single crystal with identical substitution species, global changes in ionic radius or average carrier concentration cannot account for the observed variations. Instead, local fluctuations in the strength or coherence of out-of-plane disorder scattering are more plausible. In this regime, stronger local disorder can simultaneously weaken both PG and SC correlations by reducing QP coherence, leading to the observed positive correlation. Alternatively, if residual QP states remain in the antinodal regions after the PG opens, these states may contribute to SC pairing at lower temperatures~\cite{norman1998,kondo2009}. In this view, regions with a more robust PG can naturally sustain stronger SC correlations, providing a consistent explanation for the positive correlation observed here. Because optical measurements probe averaged carrier dynamics within a finite beam size, the present observations capture mesoscale variations of electronic coherence rather than compositional differences and thus do not contradict STM or ARPES findings.

In addition to doping and disorder, other possible contributions may include local strain, which can modulate the electronic structure on comparable length scales, or competing correlations such as charge order and short-range interactions~\cite{arpaia2021charge}. These scenarios are qualitatively consistent with the micrometer-scale variations revealed by the optical imaging; direct verification will require complementary probes and lies beyond the scope of this study.

\section{Summary and Conclusions}

We have performed spatially and temporally resolved ultrafast pump-probe reflectivity spectroscopy to elucidate the interplay between pseudogap and superconducting states in optimally doped single-layer La-Bi2201. One-dimensional line scans and two-dimensional imaging revealed micrometer-scale variations in transient reflectivity governed by local differences in the threshold fluence required to disrupt either state. These threshold fluences correlate locally with the superconducting transition temperature and pseudogap energy, and their spatial variations closely track each other, highlighting a strong local link between the two states. This finding introduces a methodology complementary to momentum- and real-space probes such as ARPES and STM, and establishes a versatile framework for disentangling and correlating competing or intertwined electronic states. More broadly, the approach opens opportunities for systematic exploration of spatially inhomogeneous quantum phenomena in correlated electron systems.

Extending this threshold-based spatial analysis to other cuprates, including those with different doping levels, will be important for assessing the generality of the observed correlation and is left for future work.

\begin{acknowledgments}
We thank T. Mertelj and Y. Fukasawa for important contributions to the experiment and data analysis. Y.~T. acknowledges support from the Japan Society for the Promotion of Science (JSPS, 22H01978).
\end{acknowledgments}

 \bibliographystyle{apsrev}

\bibliography{YT202503}

\begin{thebibliography}{42}
\expandafter\ifx\csname natexlab\endcsname\relax\def\natexlab#1{#1}\fi
\expandafter\ifx\csname bibnamefont\endcsname\relax
  \def\bibnamefont#1{#1}\fi
\expandafter\ifx\csname bibfnamefont\endcsname\relax
  \def\bibfnamefont#1{#1}\fi
\expandafter\ifx\csname citenamefont\endcsname\relax
  \def\citenamefont#1{#1}\fi
\expandafter\ifx\csname url\endcsname\relax
  \def\url#1{\texttt{#1}}\fi
\expandafter\ifx\csname urlprefix\endcsname\relax\def\urlprefix{URL }\fi
\providecommand{\bibinfo}[2]{#2}
\providecommand{\eprint}[2][]{\url{#2}}

\bibitem[{\citenamefont{Timusk and Statt}(1999)}]{timusk1999}
\bibinfo{author}{\bibfnamefont{T.}~\bibnamefont{Timusk}} \bibnamefont{and}
  \bibinfo{author}{\bibfnamefont{B.}~\bibnamefont{Statt}},
  \bibinfo{journal}{Reports on Progress in Physics}
  \textbf{\bibinfo{volume}{62}}, \bibinfo{pages}{61} (\bibinfo{year}{1999}),
  \urlprefix\url{https://doi.org/10.1088/0034-4885/62/1/002}.

\bibitem[{\citenamefont{Norman et~al.}(1998)\citenamefont{Norman, Randeria,
  Ding, and Campuzano}}]{norman1998}
\bibinfo{author}{\bibfnamefont{M.~R.} \bibnamefont{Norman}},
  \bibinfo{author}{\bibfnamefont{M.}~\bibnamefont{Randeria}},
  \bibinfo{author}{\bibfnamefont{H.}~\bibnamefont{Ding}}, \bibnamefont{and}
  \bibinfo{author}{\bibfnamefont{J.~C.} \bibnamefont{Campuzano}},
  \bibinfo{journal}{Phys. Rev. B} \textbf{\bibinfo{volume}{57}},
  \bibinfo{pages}{R11093} (\bibinfo{year}{1998}),
  \urlprefix\url{https://link.aps.org/doi/10.1103/PhysRevB.57.R11093}.

\bibitem[{\citenamefont{Keimer et~al.}(2015)\citenamefont{Keimer, Kivelson,
  Norman, Uchida, and Zaanen}}]{keimer2015}
\bibinfo{author}{\bibfnamefont{B.}~\bibnamefont{Keimer}},
  \bibinfo{author}{\bibfnamefont{S.~A.} \bibnamefont{Kivelson}},
  \bibinfo{author}{\bibfnamefont{M.~R.} \bibnamefont{Norman}},
  \bibinfo{author}{\bibfnamefont{S.}~\bibnamefont{Uchida}}, \bibnamefont{and}
  \bibinfo{author}{\bibfnamefont{J.}~\bibnamefont{Zaanen}},
  \bibinfo{journal}{Nature} \textbf{\bibinfo{volume}{518}},
  \bibinfo{pages}{179} (\bibinfo{year}{2015}),
  \urlprefix\url{https://doi.org/10.1038/nature14165}.

\bibitem[{\citenamefont{Hashimoto et~al.}(2014)\citenamefont{Hashimoto, Vishik,
  He, Devereaux, and Shen}}]{hashimoto2014}
\bibinfo{author}{\bibfnamefont{M.}~\bibnamefont{Hashimoto}},
  \bibinfo{author}{\bibfnamefont{I.~M.} \bibnamefont{Vishik}},
  \bibinfo{author}{\bibfnamefont{R.-H.} \bibnamefont{He}},
  \bibinfo{author}{\bibfnamefont{T.~P.} \bibnamefont{Devereaux}},
  \bibnamefont{and} \bibinfo{author}{\bibfnamefont{Z.-X.} \bibnamefont{Shen}},
  \bibinfo{journal}{Nature Physics} \textbf{\bibinfo{volume}{10}},
  \bibinfo{pages}{483} (\bibinfo{year}{2014}),
  \urlprefix\url{https://doi.org/10.1038/nphys3009}.

\bibitem[{\citenamefont{Tanaka et~al.}(2006)\citenamefont{Tanaka, Lee, Lu,
  Fujimori, Fujii, null, Terasaki, Scalapino, Devereaux, Hussain
  et~al.}}]{tanaka2006}
\bibinfo{author}{\bibfnamefont{K.}~\bibnamefont{Tanaka}},
  \bibinfo{author}{\bibfnamefont{W.~S.} \bibnamefont{Lee}},
  \bibinfo{author}{\bibfnamefont{D.~H.} \bibnamefont{Lu}},
  \bibinfo{author}{\bibfnamefont{A.}~\bibnamefont{Fujimori}},
  \bibinfo{author}{\bibfnamefont{T.}~\bibnamefont{Fujii}},
  \bibinfo{author}{\bibnamefont{null}},
  \bibinfo{author}{\bibfnamefont{I.}~\bibnamefont{Terasaki}},
  \bibinfo{author}{\bibfnamefont{D.~J.} \bibnamefont{Scalapino}},
  \bibinfo{author}{\bibfnamefont{T.~P.} \bibnamefont{Devereaux}},
  \bibinfo{author}{\bibfnamefont{Z.}~\bibnamefont{Hussain}},
  \bibnamefont{et~al.}, \bibinfo{journal}{Science}
  \textbf{\bibinfo{volume}{314}}, \bibinfo{pages}{1910} (\bibinfo{year}{2006}),
  \urlprefix\url{https://www.science.org/doi/abs/10.1126/science.1133411}.

\bibitem[{\citenamefont{Lee et~al.}(2006)\citenamefont{Lee, Fujita, McElroy,
  Slezak, Wang, Aiura, Bando, Ishikado, Masui, Zhu et~al.}}]{lee2006}
\bibinfo{author}{\bibfnamefont{J.}~\bibnamefont{Lee}},
  \bibinfo{author}{\bibfnamefont{K.}~\bibnamefont{Fujita}},
  \bibinfo{author}{\bibfnamefont{K.}~\bibnamefont{McElroy}},
  \bibinfo{author}{\bibfnamefont{J.}~\bibnamefont{Slezak}},
  \bibinfo{author}{\bibfnamefont{M.}~\bibnamefont{Wang}},
  \bibinfo{author}{\bibfnamefont{Y.}~\bibnamefont{Aiura}},
  \bibinfo{author}{\bibfnamefont{H.}~\bibnamefont{Bando}},
  \bibinfo{author}{\bibfnamefont{M.}~\bibnamefont{Ishikado}},
  \bibinfo{author}{\bibfnamefont{T.}~\bibnamefont{Masui}},
  \bibinfo{author}{\bibfnamefont{J.-X.} \bibnamefont{Zhu}},
  \bibnamefont{et~al.}, \bibinfo{journal}{Nature}
  \textbf{\bibinfo{volume}{442}}, \bibinfo{pages}{546} (\bibinfo{year}{2006}),
  \urlprefix\url{https://www.nature.com/articles/nature04973}.

\bibitem[{\citenamefont{Kondo et~al.}(2007)\citenamefont{Kondo, Takeuchi,
  Kaminski, Tsuda, and Shin}}]{kondo2007}
\bibinfo{author}{\bibfnamefont{T.}~\bibnamefont{Kondo}},
  \bibinfo{author}{\bibfnamefont{T.}~\bibnamefont{Takeuchi}},
  \bibinfo{author}{\bibfnamefont{A.}~\bibnamefont{Kaminski}},
  \bibinfo{author}{\bibfnamefont{S.}~\bibnamefont{Tsuda}}, \bibnamefont{and}
  \bibinfo{author}{\bibfnamefont{S.}~\bibnamefont{Shin}},
  \bibinfo{journal}{Phys. Rev. Lett.} \textbf{\bibinfo{volume}{98}},
  \bibinfo{pages}{267004} (\bibinfo{year}{2007}).

\bibitem[{\citenamefont{Okada et~al.}(2008)\citenamefont{Okada, Takeuchi,
  Shimoyamada, Shin, and Ikuta}}]{okada2008}
\bibinfo{author}{\bibfnamefont{Y.}~\bibnamefont{Okada}},
  \bibinfo{author}{\bibfnamefont{T.}~\bibnamefont{Takeuchi}},
  \bibinfo{author}{\bibfnamefont{A.}~\bibnamefont{Shimoyamada}},
  \bibinfo{author}{\bibfnamefont{S.}~\bibnamefont{Shin}}, \bibnamefont{and}
  \bibinfo{author}{\bibfnamefont{H.}~\bibnamefont{Ikuta}},
  \bibinfo{journal}{Journal of Physics and Chemistry of Solids}
  \textbf{\bibinfo{volume}{69}}, \bibinfo{pages}{2989} (\bibinfo{year}{2008}).

\bibitem[{\citenamefont{Gomes et~al.}(2007)\citenamefont{Gomes, Pasupathy,
  Pushp, Ono, Ando, and Yazdani}}]{gomes2007}
\bibinfo{author}{\bibfnamefont{K.~K.} \bibnamefont{Gomes}},
  \bibinfo{author}{\bibfnamefont{A.~N.} \bibnamefont{Pasupathy}},
  \bibinfo{author}{\bibfnamefont{A.}~\bibnamefont{Pushp}},
  \bibinfo{author}{\bibfnamefont{S.}~\bibnamefont{Ono}},
  \bibinfo{author}{\bibfnamefont{Y.}~\bibnamefont{Ando}}, \bibnamefont{and}
  \bibinfo{author}{\bibfnamefont{A.}~\bibnamefont{Yazdani}},
  \bibinfo{journal}{Nature} \textbf{\bibinfo{volume}{447}},
  \bibinfo{pages}{569} (\bibinfo{year}{2007}),
  \urlprefix\url{https://doi.org/10.1038/nature05881}.

\bibitem[{\citenamefont{Vershinin et~al.}(2004)\citenamefont{Vershinin, Misra,
  Ono, Abe, Ando, and Yazdani}}]{Vershinin2004}
\bibinfo{author}{\bibfnamefont{M.}~\bibnamefont{Vershinin}},
  \bibinfo{author}{\bibfnamefont{S.}~\bibnamefont{Misra}},
  \bibinfo{author}{\bibfnamefont{S.}~\bibnamefont{Ono}},
  \bibinfo{author}{\bibfnamefont{Y.}~\bibnamefont{Abe}},
  \bibinfo{author}{\bibfnamefont{Y.}~\bibnamefont{Ando}}, \bibnamefont{and}
  \bibinfo{author}{\bibfnamefont{A.}~\bibnamefont{Yazdani}},
  \bibinfo{journal}{Science} \textbf{\bibinfo{volume}{303}},
  \bibinfo{pages}{1995} (\bibinfo{year}{2004}),
  \eprint{https://www.science.org/doi/pdf/10.1126/science.1093384},
  \urlprefix\url{https://www.science.org/doi/abs/10.1126/science.1093384}.

\bibitem[{\citenamefont{Kohsaka et~al.}(2008)\citenamefont{Kohsaka, Taylor,
  Wahl, Schmidt, Lee, Fujita, Alldredge, McElroy, Lee, Eisaki
  et~al.}}]{kohsaka2008}
\bibinfo{author}{\bibfnamefont{Y.}~\bibnamefont{Kohsaka}},
  \bibinfo{author}{\bibfnamefont{C.}~\bibnamefont{Taylor}},
  \bibinfo{author}{\bibfnamefont{P.}~\bibnamefont{Wahl}},
  \bibinfo{author}{\bibfnamefont{A.}~\bibnamefont{Schmidt}},
  \bibinfo{author}{\bibfnamefont{J.}~\bibnamefont{Lee}},
  \bibinfo{author}{\bibfnamefont{K.}~\bibnamefont{Fujita}},
  \bibinfo{author}{\bibfnamefont{J.}~\bibnamefont{Alldredge}},
  \bibinfo{author}{\bibfnamefont{K.}~\bibnamefont{McElroy}},
  \bibinfo{author}{\bibfnamefont{J.}~\bibnamefont{Lee}},
  \bibinfo{author}{\bibfnamefont{H.}~\bibnamefont{Eisaki}},
  \bibnamefont{et~al.}, \bibinfo{journal}{Nature}
  \textbf{\bibinfo{volume}{454}}, \bibinfo{pages}{1072} (\bibinfo{year}{2008}),
  \urlprefix\url{https://doi.org/10.1038/nature07243}.

\bibitem[{\citenamefont{Ideta et~al.}(2010)\citenamefont{Ideta, Takashima,
  Hashimoto, Yoshida, Fujimori, Anzai, Fujita, Nakashima, Ino, Arita
  et~al.}}]{ideta2010}
\bibinfo{author}{\bibfnamefont{S.}~\bibnamefont{Ideta}},
  \bibinfo{author}{\bibfnamefont{K.}~\bibnamefont{Takashima}},
  \bibinfo{author}{\bibfnamefont{M.}~\bibnamefont{Hashimoto}},
  \bibinfo{author}{\bibfnamefont{T.}~\bibnamefont{Yoshida}},
  \bibinfo{author}{\bibfnamefont{A.}~\bibnamefont{Fujimori}},
  \bibinfo{author}{\bibfnamefont{H.}~\bibnamefont{Anzai}},
  \bibinfo{author}{\bibfnamefont{T.}~\bibnamefont{Fujita}},
  \bibinfo{author}{\bibfnamefont{Y.}~\bibnamefont{Nakashima}},
  \bibinfo{author}{\bibfnamefont{A.}~\bibnamefont{Ino}},
  \bibinfo{author}{\bibfnamefont{M.}~\bibnamefont{Arita}},
  \bibnamefont{et~al.}, \bibinfo{journal}{Phys. Rev. Lett.}
  \textbf{\bibinfo{volume}{104}}, \bibinfo{pages}{227001}
  (\bibinfo{year}{2010}),
  \urlprefix\url{https://link.aps.org/doi/10.1103/PhysRevLett.104.227001}.

\bibitem[{\citenamefont{Tajima et~al.}(2024)\citenamefont{Tajima, Itoh,
  Mizutamari, Miyasaka, Nakajima, Sasaki, Yamaguchi, Harada, and
  Watanabe}}]{tajima2024}
\bibinfo{author}{\bibfnamefont{S.}~\bibnamefont{Tajima}},
  \bibinfo{author}{\bibfnamefont{Y.}~\bibnamefont{Itoh}},
  \bibinfo{author}{\bibfnamefont{K.}~\bibnamefont{Mizutamari}},
  \bibinfo{author}{\bibfnamefont{S.}~\bibnamefont{Miyasaka}},
  \bibinfo{author}{\bibfnamefont{M.}~\bibnamefont{Nakajima}},
  \bibinfo{author}{\bibfnamefont{N.}~\bibnamefont{Sasaki}},
  \bibinfo{author}{\bibfnamefont{S.}~\bibnamefont{Yamaguchi}},
  \bibinfo{author}{\bibfnamefont{K.-i.} \bibnamefont{Harada}},
  \bibnamefont{and} \bibinfo{author}{\bibfnamefont{T.}~\bibnamefont{Watanabe}},
  \bibinfo{journal}{Journal of the Physical Society of Japan}
  \textbf{\bibinfo{volume}{93}}, \bibinfo{pages}{103701}
  (\bibinfo{year}{2024}), \eprint{https://doi.org/10.7566/JPSJ.93.103701},
  \urlprefix\url{https://doi.org/10.7566/JPSJ.93.103701}.

\bibitem[{\citenamefont{Niu et~al.}(2024)\citenamefont{Niu, Larrazabal,
  Gozlinski, Sato, Bastiaans, Benschop, Ge, Blanter, Gu, Swart
  et~al.}}]{niu2024}
\bibinfo{author}{\bibfnamefont{J.}~\bibnamefont{Niu}},
  \bibinfo{author}{\bibfnamefont{M.~O.} \bibnamefont{Larrazabal}},
  \bibinfo{author}{\bibfnamefont{T.}~\bibnamefont{Gozlinski}},
  \bibinfo{author}{\bibfnamefont{Y.}~\bibnamefont{Sato}},
  \bibinfo{author}{\bibfnamefont{K.~M.} \bibnamefont{Bastiaans}},
  \bibinfo{author}{\bibfnamefont{T.}~\bibnamefont{Benschop}},
  \bibinfo{author}{\bibfnamefont{J.-F.} \bibnamefont{Ge}},
  \bibinfo{author}{\bibfnamefont{Y.~M.} \bibnamefont{Blanter}},
  \bibinfo{author}{\bibfnamefont{G.}~\bibnamefont{Gu}},
  \bibinfo{author}{\bibfnamefont{I.}~\bibnamefont{Swart}}, \bibnamefont{et~al.}
  (\bibinfo{year}{2024}), \eprint{2409.15928},
  \urlprefix\url{https://arxiv.org/abs/2409.15928}.

\bibitem[{\citenamefont{Ye et~al.}(2023)\citenamefont{Ye, Zou, Yan, Ji, Xu,
  Dong, Chen, Zhou, and Wang}}]{ye2023}
\bibinfo{author}{\bibfnamefont{S.}~\bibnamefont{Ye}},
  \bibinfo{author}{\bibfnamefont{C.}~\bibnamefont{Zou}},
  \bibinfo{author}{\bibfnamefont{H.}~\bibnamefont{Yan}},
  \bibinfo{author}{\bibfnamefont{Y.}~\bibnamefont{Ji}},
  \bibinfo{author}{\bibfnamefont{M.}~\bibnamefont{Xu}},
  \bibinfo{author}{\bibfnamefont{Z.}~\bibnamefont{Dong}},
  \bibinfo{author}{\bibfnamefont{Y.}~\bibnamefont{Chen}},
  \bibinfo{author}{\bibfnamefont{X.}~\bibnamefont{Zhou}}, \bibnamefont{and}
  \bibinfo{author}{\bibfnamefont{Y.}~\bibnamefont{Wang}},
  \bibinfo{journal}{Nature Physics} \textbf{\bibinfo{volume}{19}},
  \bibinfo{pages}{1301} (\bibinfo{year}{2023}),
  \urlprefix\url{https://doi.org/10.1038/s41567-023-02100-9}.

\bibitem[{\citenamefont{de~la Torre et~al.}(2021)\citenamefont{de~la Torre,
  Kennes, Claassen, Gerber, McIver, and Sentef}}]{de2021}
\bibinfo{author}{\bibfnamefont{A.}~\bibnamefont{de~la Torre}},
  \bibinfo{author}{\bibfnamefont{D.~M.} \bibnamefont{Kennes}},
  \bibinfo{author}{\bibfnamefont{M.}~\bibnamefont{Claassen}},
  \bibinfo{author}{\bibfnamefont{S.}~\bibnamefont{Gerber}},
  \bibinfo{author}{\bibfnamefont{J.~W.} \bibnamefont{McIver}},
  \bibnamefont{and} \bibinfo{author}{\bibfnamefont{M.~A.}
  \bibnamefont{Sentef}}, \bibinfo{journal}{Rev. Mod. Phys.}
  \textbf{\bibinfo{volume}{93}}, \bibinfo{pages}{041002}
  (\bibinfo{year}{2021}),
  \urlprefix\url{https://link.aps.org/doi/10.1103/RevModPhys.93.041002}.

\bibitem[{\citenamefont{Demsar et~al.}(1999)\citenamefont{Demsar, Podobnik,
  Kabanov, Wolf, and Mihailovic}}]{demsar1999}
\bibinfo{author}{\bibfnamefont{J.}~\bibnamefont{Demsar}},
  \bibinfo{author}{\bibfnamefont{B.}~\bibnamefont{Podobnik}},
  \bibinfo{author}{\bibfnamefont{V.~V.} \bibnamefont{Kabanov}},
  \bibinfo{author}{\bibfnamefont{T.}~\bibnamefont{Wolf}}, \bibnamefont{and}
  \bibinfo{author}{\bibfnamefont{D.}~\bibnamefont{Mihailovic}},
  \bibinfo{journal}{Phys. Rev. Lett.} \textbf{\bibinfo{volume}{82}},
  \bibinfo{pages}{4918} (\bibinfo{year}{1999}),
  \urlprefix\url{https://link.aps.org/doi/10.1103/PhysRevLett.82.4918}.

\bibitem[{\citenamefont{Kaindl et~al.}(2000)\citenamefont{Kaindl, Woerner,
  Elsaesser, Smith, Ryan, Farnan, McCurry, and Walmsley}}]{kaindl2000}
\bibinfo{author}{\bibfnamefont{R.~A.} \bibnamefont{Kaindl}},
  \bibinfo{author}{\bibfnamefont{M.}~\bibnamefont{Woerner}},
  \bibinfo{author}{\bibfnamefont{T.}~\bibnamefont{Elsaesser}},
  \bibinfo{author}{\bibfnamefont{D.~C.} \bibnamefont{Smith}},
  \bibinfo{author}{\bibfnamefont{J.~F.} \bibnamefont{Ryan}},
  \bibinfo{author}{\bibfnamefont{G.~A.} \bibnamefont{Farnan}},
  \bibinfo{author}{\bibfnamefont{M.~P.} \bibnamefont{McCurry}},
  \bibnamefont{and} \bibinfo{author}{\bibfnamefont{D.~G.}
  \bibnamefont{Walmsley}}, \bibinfo{journal}{Science}
  \textbf{\bibinfo{volume}{287}}, \bibinfo{pages}{470} (\bibinfo{year}{2000}),
  \urlprefix\url{https://www.science.org/doi/abs/10.1126/science.287.5452.470}.

\bibitem[{\citenamefont{Luo et~al.}(2006)\citenamefont{Luo, Hsieh, Chen, Shih,
  Chen, Wu, Juang, Lin, Uen, and Gou}}]{luo2006}
\bibinfo{author}{\bibfnamefont{C.~W.} \bibnamefont{Luo}},
  \bibinfo{author}{\bibfnamefont{C.~C.} \bibnamefont{Hsieh}},
  \bibinfo{author}{\bibfnamefont{Y.-J.} \bibnamefont{Chen}},
  \bibinfo{author}{\bibfnamefont{P.~T.} \bibnamefont{Shih}},
  \bibinfo{author}{\bibfnamefont{M.~H.} \bibnamefont{Chen}},
  \bibinfo{author}{\bibfnamefont{K.~H.} \bibnamefont{Wu}},
  \bibinfo{author}{\bibfnamefont{J.~Y.} \bibnamefont{Juang}},
  \bibinfo{author}{\bibfnamefont{J.-Y.} \bibnamefont{Lin}},
  \bibinfo{author}{\bibfnamefont{T.~M.} \bibnamefont{Uen}}, \bibnamefont{and}
  \bibinfo{author}{\bibfnamefont{Y.~S.} \bibnamefont{Gou}},
  \bibinfo{journal}{Phys. Rev. B} \textbf{\bibinfo{volume}{74}},
  \bibinfo{pages}{184525} (\bibinfo{year}{2006}),
  \urlprefix\url{https://link.aps.org/doi/10.1103/PhysRevB.74.184525}.

\bibitem[{\citenamefont{Toda et~al.}(2011)\citenamefont{Toda, Mertelj, Kusar,
  Kurosawa, Oda, Ido, and Mihailovic}}]{toda2011}
\bibinfo{author}{\bibfnamefont{Y.}~\bibnamefont{Toda}},
  \bibinfo{author}{\bibfnamefont{T.}~\bibnamefont{Mertelj}},
  \bibinfo{author}{\bibfnamefont{P.}~\bibnamefont{Kusar}},
  \bibinfo{author}{\bibfnamefont{T.}~\bibnamefont{Kurosawa}},
  \bibinfo{author}{\bibfnamefont{M.}~\bibnamefont{Oda}},
  \bibinfo{author}{\bibfnamefont{M.}~\bibnamefont{Ido}}, \bibnamefont{and}
  \bibinfo{author}{\bibfnamefont{D.}~\bibnamefont{Mihailovic}},
  \bibinfo{journal}{Physical Review B} \textbf{\bibinfo{volume}{84}},
  \bibinfo{pages}{174516} (\bibinfo{year}{2011}).

\bibitem[{\citenamefont{Coslovich et~al.}(2013)\citenamefont{Coslovich,
  Giannetti, Cilento, Dal~Conte, Abebaw, Bossini, Ferrini, Eisaki, Greven,
  Damascelli et~al.}}]{coslovich2013}
\bibinfo{author}{\bibfnamefont{G.}~\bibnamefont{Coslovich}},
  \bibinfo{author}{\bibfnamefont{C.}~\bibnamefont{Giannetti}},
  \bibinfo{author}{\bibfnamefont{F.}~\bibnamefont{Cilento}},
  \bibinfo{author}{\bibfnamefont{S.}~\bibnamefont{Dal~Conte}},
  \bibinfo{author}{\bibfnamefont{T.}~\bibnamefont{Abebaw}},
  \bibinfo{author}{\bibfnamefont{D.}~\bibnamefont{Bossini}},
  \bibinfo{author}{\bibfnamefont{G.}~\bibnamefont{Ferrini}},
  \bibinfo{author}{\bibfnamefont{H.}~\bibnamefont{Eisaki}},
  \bibinfo{author}{\bibfnamefont{M.}~\bibnamefont{Greven}},
  \bibinfo{author}{\bibfnamefont{A.}~\bibnamefont{Damascelli}},
  \bibnamefont{et~al.}, \bibinfo{journal}{Physical Review Letters}
  \textbf{\bibinfo{volume}{110}}, \bibinfo{pages}{107003}
  (\bibinfo{year}{2013}).

\bibitem[{\citenamefont{Toda et~al.}(2023)\citenamefont{Toda, Tsuchiya, Yamane,
  Morita, Oda, Kurosawa, Mertelj, and Mihailovic}}]{toda2023}
\bibinfo{author}{\bibfnamefont{Y.}~\bibnamefont{Toda}},
  \bibinfo{author}{\bibfnamefont{S.}~\bibnamefont{Tsuchiya}},
  \bibinfo{author}{\bibfnamefont{K.}~\bibnamefont{Yamane}},
  \bibinfo{author}{\bibfnamefont{R.}~\bibnamefont{Morita}},
  \bibinfo{author}{\bibfnamefont{M.}~\bibnamefont{Oda}},
  \bibinfo{author}{\bibfnamefont{T.}~\bibnamefont{Kurosawa}},
  \bibinfo{author}{\bibfnamefont{T.}~\bibnamefont{Mertelj}}, \bibnamefont{and}
  \bibinfo{author}{\bibfnamefont{D.}~\bibnamefont{Mihailovic}},
  \bibinfo{journal}{Opt. Express} \textbf{\bibinfo{volume}{31}},
  \bibinfo{pages}{17537} (\bibinfo{year}{2023}),
  \urlprefix\url{https://opg.optica.org/oe/abstract.cfm?URI=oe-31-11-17537}.

\bibitem[{\citenamefont{Akiba et~al.}(2024)\citenamefont{Akiba, Toda, Tsuchiya,
  Oda, Kurosawa, Mihailovic, and Mertelj}}]{akiba2024}
\bibinfo{author}{\bibfnamefont{T.}~\bibnamefont{Akiba}},
  \bibinfo{author}{\bibfnamefont{Y.}~\bibnamefont{Toda}},
  \bibinfo{author}{\bibfnamefont{S.}~\bibnamefont{Tsuchiya}},
  \bibinfo{author}{\bibfnamefont{M.}~\bibnamefont{Oda}},
  \bibinfo{author}{\bibfnamefont{T.}~\bibnamefont{Kurosawa}},
  \bibinfo{author}{\bibfnamefont{D.}~\bibnamefont{Mihailovic}},
  \bibnamefont{and} \bibinfo{author}{\bibfnamefont{T.}~\bibnamefont{Mertelj}},
  \bibinfo{journal}{Phys. Rev. B} \textbf{\bibinfo{volume}{109}},
  \bibinfo{pages}{014503} (\bibinfo{year}{2024}),
  \urlprefix\url{https://link.aps.org/doi/10.1103/PhysRevB.109.014503}.

\bibitem[{\citenamefont{Eisaki et~al.}(2004)\citenamefont{Eisaki, Kaneko, Feng,
  Damascelli, Mang, Shen, Shen, and Greven}}]{eisaki2004}
\bibinfo{author}{\bibfnamefont{H.}~\bibnamefont{Eisaki}},
  \bibinfo{author}{\bibfnamefont{N.}~\bibnamefont{Kaneko}},
  \bibinfo{author}{\bibfnamefont{D.~L.} \bibnamefont{Feng}},
  \bibinfo{author}{\bibfnamefont{A.}~\bibnamefont{Damascelli}},
  \bibinfo{author}{\bibfnamefont{P.~K.} \bibnamefont{Mang}},
  \bibinfo{author}{\bibfnamefont{K.~M.} \bibnamefont{Shen}},
  \bibinfo{author}{\bibfnamefont{Z.-X.} \bibnamefont{Shen}}, \bibnamefont{and}
  \bibinfo{author}{\bibfnamefont{M.}~\bibnamefont{Greven}},
  \bibinfo{journal}{Phys. Rev. B} \textbf{\bibinfo{volume}{69}},
  \bibinfo{pages}{064512} (\bibinfo{year}{2004}),
  \urlprefix\url{https://link.aps.org/doi/10.1103/PhysRevB.69.064512}.

\bibitem[{\citenamefont{Kurosawa et~al.}(2010)\citenamefont{Kurosawa, Yoneyama,
  Takano, Hagiwara, Inoue, Hagiwara, Kurusu, Takeyama, Momono, Oda
  et~al.}}]{kurosawa2010}
\bibinfo{author}{\bibfnamefont{T.}~\bibnamefont{Kurosawa}},
  \bibinfo{author}{\bibfnamefont{T.}~\bibnamefont{Yoneyama}},
  \bibinfo{author}{\bibfnamefont{Y.}~\bibnamefont{Takano}},
  \bibinfo{author}{\bibfnamefont{M.}~\bibnamefont{Hagiwara}},
  \bibinfo{author}{\bibfnamefont{R.}~\bibnamefont{Inoue}},
  \bibinfo{author}{\bibfnamefont{N.}~\bibnamefont{Hagiwara}},
  \bibinfo{author}{\bibfnamefont{K.}~\bibnamefont{Kurusu}},
  \bibinfo{author}{\bibfnamefont{K.}~\bibnamefont{Takeyama}},
  \bibinfo{author}{\bibfnamefont{N.}~\bibnamefont{Momono}},
  \bibinfo{author}{\bibfnamefont{M.}~\bibnamefont{Oda}}, \bibnamefont{et~al.},
  \bibinfo{journal}{Phys. Rev. B} \textbf{\bibinfo{volume}{81}},
  \bibinfo{pages}{094519} (\bibinfo{year}{2010}),
  \urlprefix\url{https://link.aps.org/doi/10.1103/PhysRevB.81.094519}.

\bibitem[{\citenamefont{Kurosawa et~al.}(2016)\citenamefont{Kurosawa, Takeyama,
  Baar, Shibata, Kataoka, Mizuta, Yoshida, Momono, Oda, and
  Ido}}]{kurosawa2016}
\bibinfo{author}{\bibfnamefont{T.}~\bibnamefont{Kurosawa}},
  \bibinfo{author}{\bibfnamefont{K.}~\bibnamefont{Takeyama}},
  \bibinfo{author}{\bibfnamefont{S.}~\bibnamefont{Baar}},
  \bibinfo{author}{\bibfnamefont{Y.}~\bibnamefont{Shibata}},
  \bibinfo{author}{\bibfnamefont{M.}~\bibnamefont{Kataoka}},
  \bibinfo{author}{\bibfnamefont{S.}~\bibnamefont{Mizuta}},
  \bibinfo{author}{\bibfnamefont{H.}~\bibnamefont{Yoshida}},
  \bibinfo{author}{\bibfnamefont{N.}~\bibnamefont{Momono}},
  \bibinfo{author}{\bibfnamefont{M.}~\bibnamefont{Oda}}, \bibnamefont{and}
  \bibinfo{author}{\bibfnamefont{M.}~\bibnamefont{Ido}},
  \bibinfo{journal}{Journal of the Physical Society of Japan}
  \textbf{\bibinfo{volume}{85}}, \bibinfo{pages}{044709}
  (\bibinfo{year}{2016}).

\bibitem[{\citenamefont{Dvorsek et~al.}(2002)\citenamefont{Dvorsek, Kabanov,
  Demsar, Kazakov, Karpinski, and Mihailovic}}]{Dvorsek2002}
\bibinfo{author}{\bibfnamefont{D.}~\bibnamefont{Dvorsek}},
  \bibinfo{author}{\bibfnamefont{V.~V.} \bibnamefont{Kabanov}},
  \bibinfo{author}{\bibfnamefont{J.}~\bibnamefont{Demsar}},
  \bibinfo{author}{\bibfnamefont{S.~M.} \bibnamefont{Kazakov}},
  \bibinfo{author}{\bibfnamefont{J.}~\bibnamefont{Karpinski}},
  \bibnamefont{and}
  \bibinfo{author}{\bibfnamefont{D.}~\bibnamefont{Mihailovic}},
  \bibinfo{journal}{Phys. Rev. B} \textbf{\bibinfo{volume}{66}},
  \bibinfo{pages}{020510} (\bibinfo{year}{2002}),
  \urlprefix\url{https://link.aps.org/doi/10.1103/PhysRevB.66.020510}.

\bibitem[{\citenamefont{Segre et~al.}(2002)\citenamefont{Segre, Gedik,
  Orenstein, Bonn, Liang, and Hardy}}]{Segre2002}
\bibinfo{author}{\bibfnamefont{G.~P.} \bibnamefont{Segre}},
  \bibinfo{author}{\bibfnamefont{N.}~\bibnamefont{Gedik}},
  \bibinfo{author}{\bibfnamefont{J.}~\bibnamefont{Orenstein}},
  \bibinfo{author}{\bibfnamefont{D.~A.} \bibnamefont{Bonn}},
  \bibinfo{author}{\bibfnamefont{R.}~\bibnamefont{Liang}}, \bibnamefont{and}
  \bibinfo{author}{\bibfnamefont{W.~N.} \bibnamefont{Hardy}},
  \bibinfo{journal}{Phys. Rev. Lett.} \textbf{\bibinfo{volume}{88}},
  \bibinfo{pages}{137001} (\bibinfo{year}{2002}),
  \urlprefix\url{https://link.aps.org/doi/10.1103/PhysRevLett.88.137001}.

\bibitem[{\citenamefont{Kusar et~al.}(2005)\citenamefont{Kusar, Demsar,
  Mihailovic, and Sugai}}]{Kusar2005}
\bibinfo{author}{\bibfnamefont{P.}~\bibnamefont{Kusar}},
  \bibinfo{author}{\bibfnamefont{J.}~\bibnamefont{Demsar}},
  \bibinfo{author}{\bibfnamefont{D.}~\bibnamefont{Mihailovic}},
  \bibnamefont{and} \bibinfo{author}{\bibfnamefont{S.}~\bibnamefont{Sugai}},
  \bibinfo{journal}{Phys. Rev. B} \textbf{\bibinfo{volume}{72}},
  \bibinfo{pages}{014544} (\bibinfo{year}{2005}),
  \urlprefix\url{https://link.aps.org/doi/10.1103/PhysRevB.72.014544}.

\bibitem[{\citenamefont{Toda et~al.}(2021)\citenamefont{Toda, Tsuchiya, Oda,
  Kurosawa, Katsumata, Naseska, Mertelj, and Mihailovic}}]{toda2021}
\bibinfo{author}{\bibfnamefont{Y.}~\bibnamefont{Toda}},
  \bibinfo{author}{\bibfnamefont{S.}~\bibnamefont{Tsuchiya}},
  \bibinfo{author}{\bibfnamefont{M.}~\bibnamefont{Oda}},
  \bibinfo{author}{\bibfnamefont{T.}~\bibnamefont{Kurosawa}},
  \bibinfo{author}{\bibfnamefont{S.}~\bibnamefont{Katsumata}},
  \bibinfo{author}{\bibfnamefont{M.}~\bibnamefont{Naseska}},
  \bibinfo{author}{\bibfnamefont{T.}~\bibnamefont{Mertelj}}, \bibnamefont{and}
  \bibinfo{author}{\bibfnamefont{D.}~\bibnamefont{Mihailovic}},
  \bibinfo{journal}{Phys. Rev. B} \textbf{\bibinfo{volume}{104}},
  \bibinfo{pages}{094507} (\bibinfo{year}{2021}),
  \urlprefix\url{https://link.aps.org/doi/10.1103/PhysRevB.104.094507}.

\bibitem[{\citenamefont{Kusar et~al.}(2008)\citenamefont{Kusar, Kabanov, Sugai,
  Demsar, Mertelj, and Mihailovic}}]{kusar2008}
\bibinfo{author}{\bibfnamefont{P.}~\bibnamefont{Kusar}},
  \bibinfo{author}{\bibfnamefont{V.~V.} \bibnamefont{Kabanov}},
  \bibinfo{author}{\bibfnamefont{S.}~\bibnamefont{Sugai}},
  \bibinfo{author}{\bibfnamefont{J.}~\bibnamefont{Demsar}},
  \bibinfo{author}{\bibfnamefont{T.}~\bibnamefont{Mertelj}}, \bibnamefont{and}
  \bibinfo{author}{\bibfnamefont{D.}~\bibnamefont{Mihailovic}},
  \bibinfo{journal}{Phys. Rev. Lett.} \textbf{\bibinfo{volume}{101}},
  \bibinfo{pages}{227001} (\bibinfo{year}{2008}),
  \urlprefix\url{https://link.aps.org/doi/10.1103/PhysRevLett.101.227001}.

\bibitem[{\citenamefont{Naseska et~al.}(2018)\citenamefont{Naseska, Pogrebna,
  Cao, Xu, Mihailovic, and Mertelj}}]{naseska2018}
\bibinfo{author}{\bibfnamefont{M.}~\bibnamefont{Naseska}},
  \bibinfo{author}{\bibfnamefont{A.}~\bibnamefont{Pogrebna}},
  \bibinfo{author}{\bibfnamefont{G.}~\bibnamefont{Cao}},
  \bibinfo{author}{\bibfnamefont{Z.~A.} \bibnamefont{Xu}},
  \bibinfo{author}{\bibfnamefont{D.}~\bibnamefont{Mihailovic}},
  \bibnamefont{and} \bibinfo{author}{\bibfnamefont{T.}~\bibnamefont{Mertelj}},
  \bibinfo{journal}{Phys. Rev. B} \textbf{\bibinfo{volume}{98}},
  \bibinfo{pages}{035148} (\bibinfo{year}{2018}),
  \urlprefix\url{https://link.aps.org/doi/10.1103/PhysRevB.98.035148}.

\bibitem[{\citenamefont{Mertelj et~al.}(2009)\citenamefont{Mertelj, Kabanov,
  Gadermaier, Zhigadlo, Katrych, Karpinski, and
  Mihailovic}}]{mertelj2009distinct}
\bibinfo{author}{\bibfnamefont{T.}~\bibnamefont{Mertelj}},
  \bibinfo{author}{\bibfnamefont{V.~V.} \bibnamefont{Kabanov}},
  \bibinfo{author}{\bibfnamefont{C.}~\bibnamefont{Gadermaier}},
  \bibinfo{author}{\bibfnamefont{N.~D.} \bibnamefont{Zhigadlo}},
  \bibinfo{author}{\bibfnamefont{S.}~\bibnamefont{Katrych}},
  \bibinfo{author}{\bibfnamefont{J.}~\bibnamefont{Karpinski}},
  \bibnamefont{and}
  \bibinfo{author}{\bibfnamefont{D.}~\bibnamefont{Mihailovic}},
  \bibinfo{journal}{Physical Review Letters} \textbf{\bibinfo{volume}{102}},
  \bibinfo{pages}{117002} (\bibinfo{year}{2009}), ISSN
  \bibinfo{issn}{0031-9007, 1079-7114},
  \urlprefix\url{https://link.aps.org/doi/10.1103/PhysRevLett.102.117002}.

\bibitem[{\citenamefont{Kabanov et~al.}(1999)\citenamefont{Kabanov, Demsar,
  Podobnik, and Mihailovic}}]{kabanov1999}
\bibinfo{author}{\bibfnamefont{V.~V.} \bibnamefont{Kabanov}},
  \bibinfo{author}{\bibfnamefont{J.}~\bibnamefont{Demsar}},
  \bibinfo{author}{\bibfnamefont{B.}~\bibnamefont{Podobnik}}, \bibnamefont{and}
  \bibinfo{author}{\bibfnamefont{D.}~\bibnamefont{Mihailovic}},
  \bibinfo{journal}{Phys. Rev. B} \textbf{\bibinfo{volume}{59}},
  \bibinfo{pages}{1497} (\bibinfo{year}{1999}),
  \urlprefix\url{https://link.aps.org/doi/10.1103/PhysRevB.59.1497}.

\bibitem[{\citenamefont{Stojchevska et~al.}(2011)\citenamefont{Stojchevska,
  Kusar, Mertelj, Kabanov, Toda, Yao, and Mihailovic}}]{stojchevska2011}
\bibinfo{author}{\bibfnamefont{L.}~\bibnamefont{Stojchevska}},
  \bibinfo{author}{\bibfnamefont{P.}~\bibnamefont{Kusar}},
  \bibinfo{author}{\bibfnamefont{T.}~\bibnamefont{Mertelj}},
  \bibinfo{author}{\bibfnamefont{V.~V.} \bibnamefont{Kabanov}},
  \bibinfo{author}{\bibfnamefont{Y.}~\bibnamefont{Toda}},
  \bibinfo{author}{\bibfnamefont{X.}~\bibnamefont{Yao}}, \bibnamefont{and}
  \bibinfo{author}{\bibfnamefont{D.}~\bibnamefont{Mihailovic}},
  \bibinfo{journal}{Phys. Rev. B} \textbf{\bibinfo{volume}{84}},
  \bibinfo{pages}{180507} (\bibinfo{year}{2011}),
  \urlprefix\url{https://link.aps.org/doi/10.1103/PhysRevB.84.180507}.

\bibitem[{\citenamefont{Madan et~al.}(2015)\citenamefont{Madan, Kurosawa, Toda,
  Oda, Mertelj, and Mihailovic}}]{madan2015}
\bibinfo{author}{\bibfnamefont{I.}~\bibnamefont{Madan}},
  \bibinfo{author}{\bibfnamefont{T.}~\bibnamefont{Kurosawa}},
  \bibinfo{author}{\bibfnamefont{Y.}~\bibnamefont{Toda}},
  \bibinfo{author}{\bibfnamefont{M.}~\bibnamefont{Oda}},
  \bibinfo{author}{\bibfnamefont{T.}~\bibnamefont{Mertelj}}, \bibnamefont{and}
  \bibinfo{author}{\bibfnamefont{D.}~\bibnamefont{Mihailovic}},
  \bibinfo{journal}{Nature Communications} \textbf{\bibinfo{volume}{6}},
  \bibinfo{pages}{1} (\bibinfo{year}{2015}).

\bibitem[{\citenamefont{Kondo et~al.}(2009)\citenamefont{Kondo, Khasanov,
  Takeuchi, Schmalian, and Kaminski}}]{kondo2009}
\bibinfo{author}{\bibfnamefont{T.}~\bibnamefont{Kondo}},
  \bibinfo{author}{\bibfnamefont{R.}~\bibnamefont{Khasanov}},
  \bibinfo{author}{\bibfnamefont{T.}~\bibnamefont{Takeuchi}},
  \bibinfo{author}{\bibfnamefont{J.}~\bibnamefont{Schmalian}},
  \bibnamefont{and} \bibinfo{author}{\bibfnamefont{A.}~\bibnamefont{Kaminski}},
  \bibinfo{journal}{Nature} \textbf{\bibinfo{volume}{457}},
  \bibinfo{pages}{296} (\bibinfo{year}{2009}),
  \urlprefix\url{https://doi.org/10.1038/nature07644}.

\bibitem[{\citenamefont{Okada et~al.}(2011)\citenamefont{Okada, Kawaguchi,
  Ohkawa, Ishizaka, Takeuchi, Shin, and Ikuta}}]{okada2011}
\bibinfo{author}{\bibfnamefont{Y.}~\bibnamefont{Okada}},
  \bibinfo{author}{\bibfnamefont{T.}~\bibnamefont{Kawaguchi}},
  \bibinfo{author}{\bibfnamefont{M.}~\bibnamefont{Ohkawa}},
  \bibinfo{author}{\bibfnamefont{K.}~\bibnamefont{Ishizaka}},
  \bibinfo{author}{\bibfnamefont{T.}~\bibnamefont{Takeuchi}},
  \bibinfo{author}{\bibfnamefont{S.}~\bibnamefont{Shin}}, \bibnamefont{and}
  \bibinfo{author}{\bibfnamefont{H.}~\bibnamefont{Ikuta}},
  \bibinfo{journal}{Phys. Rev. B} \textbf{\bibinfo{volume}{83}},
  \bibinfo{pages}{104502} (\bibinfo{year}{2011}).

\bibitem[{\citenamefont{Arpaia and Ghiringhelli}(2021)}]{arpaia2021charge}
\bibinfo{author}{\bibfnamefont{R.}~\bibnamefont{Arpaia}} \bibnamefont{and}
  \bibinfo{author}{\bibfnamefont{G.}~\bibnamefont{Ghiringhelli}},
  \bibinfo{journal}{Journal of the Physical Society of Japan}
  \textbf{\bibinfo{volume}{90}}, \bibinfo{pages}{111005}
  (\bibinfo{year}{2021}),
  \urlprefix\url{https://doi.org/10.7566/JPSJ.90.111005}.

\bibitem[{\citenamefont{Liu et~al.}(2008)\citenamefont{Liu, Toda, Shimatake,
  Momono, Oda, and Ido}}]{liu2008}
\bibinfo{author}{\bibfnamefont{Y.~H.} \bibnamefont{Liu}},
  \bibinfo{author}{\bibfnamefont{Y.}~\bibnamefont{Toda}},
  \bibinfo{author}{\bibfnamefont{K.}~\bibnamefont{Shimatake}},
  \bibinfo{author}{\bibfnamefont{N.}~\bibnamefont{Momono}},
  \bibinfo{author}{\bibfnamefont{M.}~\bibnamefont{Oda}}, \bibnamefont{and}
  \bibinfo{author}{\bibfnamefont{M.}~\bibnamefont{Ido}},
  \bibinfo{journal}{Phys. Rev. Lett.} \textbf{\bibinfo{volume}{101}},
  \bibinfo{pages}{137003} (\bibinfo{year}{2008}),
  \urlprefix\url{https://link.aps.org/doi/10.1103/PhysRevLett.101.137003}.

\bibitem[{\citenamefont{Giannetti et~al.}(2011)\citenamefont{Giannetti,
  Cilento, Dal~Conte, Coslovich, Ferrini, Molegraaf, Raichle, Liang, Eisaki,
  Greven et~al.}}]{giannetti2011}
\bibinfo{author}{\bibfnamefont{C.}~\bibnamefont{Giannetti}},
  \bibinfo{author}{\bibfnamefont{F.}~\bibnamefont{Cilento}},
  \bibinfo{author}{\bibfnamefont{S.}~\bibnamefont{Dal~Conte}},
  \bibinfo{author}{\bibfnamefont{G.}~\bibnamefont{Coslovich}},
  \bibinfo{author}{\bibfnamefont{G.}~\bibnamefont{Ferrini}},
  \bibinfo{author}{\bibfnamefont{H.}~\bibnamefont{Molegraaf}},
  \bibinfo{author}{\bibfnamefont{M.}~\bibnamefont{Raichle}},
  \bibinfo{author}{\bibfnamefont{R.}~\bibnamefont{Liang}},
  \bibinfo{author}{\bibfnamefont{H.}~\bibnamefont{Eisaki}},
  \bibinfo{author}{\bibfnamefont{M.}~\bibnamefont{Greven}},
  \bibnamefont{et~al.}, \bibinfo{journal}{Nature Communications}
  \textbf{\bibinfo{volume}{2}}, \bibinfo{pages}{353} (\bibinfo{year}{2011}).

\bibitem[{\citenamefont{Toda et~al.}(2014)\citenamefont{Toda, Kawanokami,
  Kurosawa, Oda, Madan, Mertelj, Kabanov, and Mihailovic}}]{toda2014}
\bibinfo{author}{\bibfnamefont{Y.}~\bibnamefont{Toda}},
  \bibinfo{author}{\bibfnamefont{F.}~\bibnamefont{Kawanokami}},
  \bibinfo{author}{\bibfnamefont{T.}~\bibnamefont{Kurosawa}},
  \bibinfo{author}{\bibfnamefont{M.}~\bibnamefont{Oda}},
  \bibinfo{author}{\bibfnamefont{I.}~\bibnamefont{Madan}},
  \bibinfo{author}{\bibfnamefont{T.}~\bibnamefont{Mertelj}},
  \bibinfo{author}{\bibfnamefont{V.~V.} \bibnamefont{Kabanov}},
  \bibnamefont{and}
  \bibinfo{author}{\bibfnamefont{D.}~\bibnamefont{Mihailovic}},
  \bibinfo{journal}{Phys. Rev. B} \textbf{\bibinfo{volume}{90}},
  \bibinfo{pages}{094513} (\bibinfo{year}{2014}).

\end{thebibliography}

\appendix

\renewcommand{\thefigure}{S\arabic{figure}}
\renewcommand{\thetable}{S\arabic{table}}
\renewcommand{\theequation}{S\arabic{equation}}

\section{One-dimensional (1D) line scans of $\Delta R/R$}

\begin{figure}[htbp]
\centering
\includegraphics[width=0.9\columnwidth]{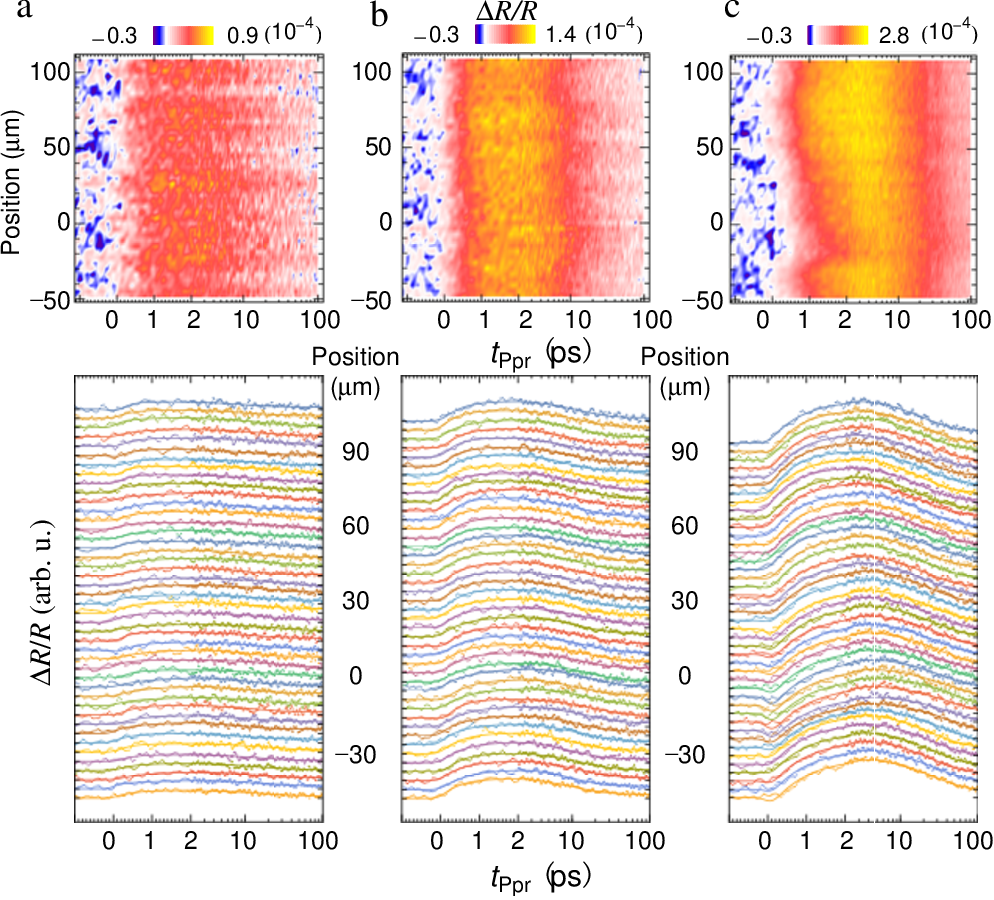}
\caption{(Color online) 1D line scans of $\Delta R/R$ obtained at $T=10~\mathrm{K}$ along the $\sim$160~$\mu$m path indicated by the dashed line in Fig.~\ref{fig_2D}(a) of the main text. Pump fluences: (a) ${\mathcal{F}}=0.6~\mu$J/cm$^2$, (b) $2.0~\mu$J/cm$^2$, (c) $20~\mu$J/cm$^2$. Upper panels: color density plots. Lower panels: corresponding $\Delta R/R$ traces. Single-exponential fits are shown as solid lines.}
\label{fig_1DSC}
\end{figure}

One-dimensional line-scan measurements of $\Delta R/R$ in La-Bi2201 at $T=10~\mathrm{K}$ under different pump fluences are shown in Fig.~\ref{fig_1DSC}. Along the dashed line indicated in Fig.~\ref{fig_2D}(a) of the main text, $\Delta R/R$ was measured at 43 positions with spacing $\Delta x = 3.75~\mu$m (total path length $\sim$160~$\mu$m) and is displayed as (upper) color density plots and (lower) transient traces. In the lower panels, the traces are vertically offset according to position. The $\Delta R/R$ response is primarily dominated by the superconducting component, characterized by a slow relaxation. From each transient trace, the superconducting response amplitude $A_{\rm SC}$ is defined as $\langle \Delta R/R \rangle_{2-10~\mathrm{ps}}$ and is plotted in Fig.~\ref{fig_1D} of the main text. As seen in the color maps, the spatial intensity distribution remains weak under low excitation [Fig.~\ref{fig_1DSC}(a)], whereas a pronounced distribution emerges under high-fluence excitation [Fig.~\ref{fig_1DSC}(c)], consistent with Figs.~\ref{fig_2D}(c),(d) of the main text.

\begin{figure}[htbp]
\centering
\includegraphics[width=0.8\columnwidth]{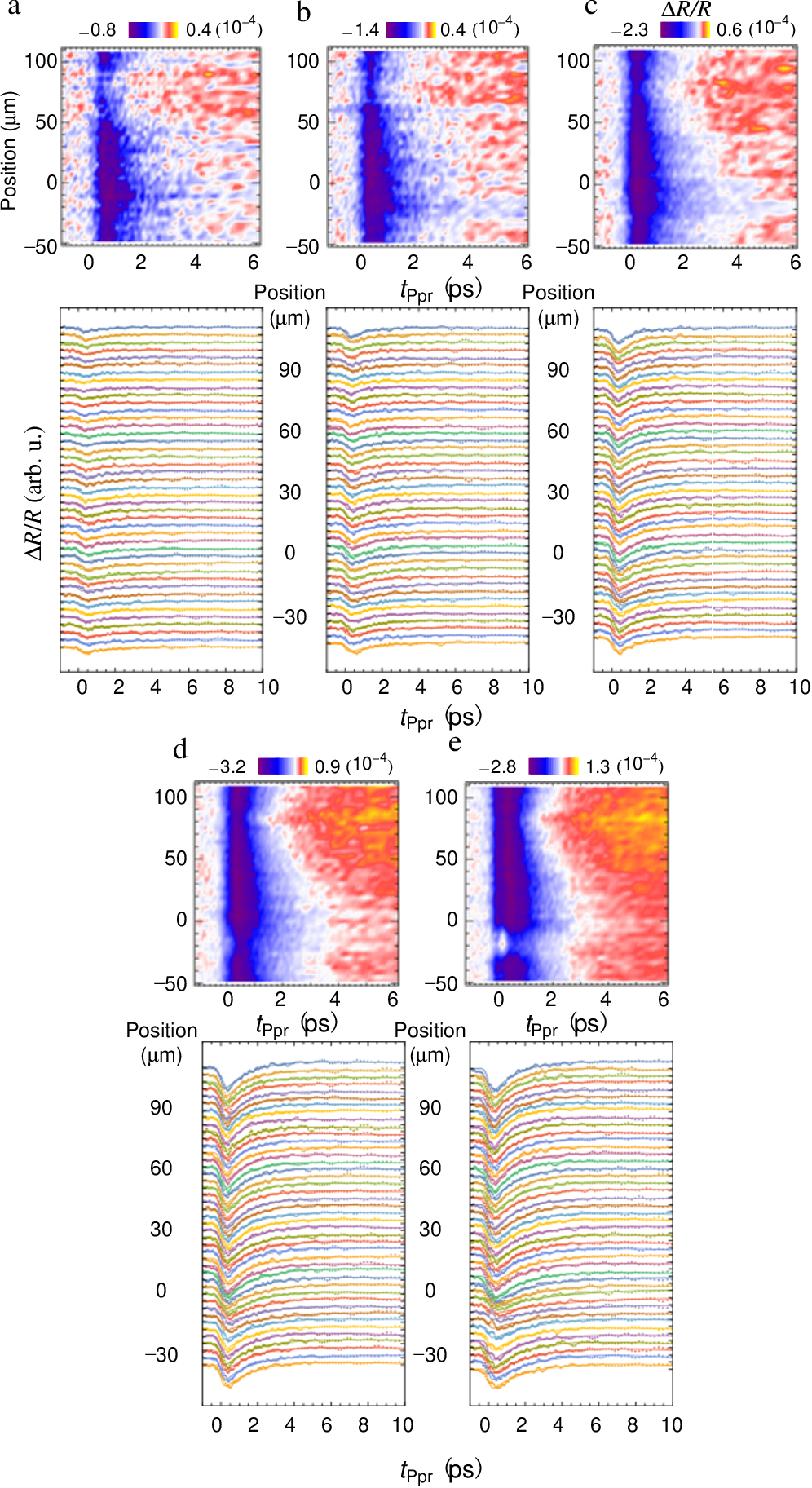}
\caption{(Color online) Same as Fig.~\ref{fig_1DSC}, but measured at $T=50~\mathrm{K}$. Pump fluences: (a) $5.2~\mu$J/cm$^2$, (b) $9.4~\mu$J/cm$^2$, (c) $20~\mu$J/cm$^2$, (d) $52~\mu$J/cm$^2$, and (e) $104~\mu$J/cm$^2$.}
\label{fig_1DPG}
\end{figure}

Figure~\ref{fig_1DPG} presents analogous line-scan measurements at $T=50~\mathrm{K}$. The fast-decaying negative $\Delta R/R$ corresponds to the PG response. To emphasize this component, the time window of the horizontal axis is set to $t_{\rm Ppr}=-1$ to 6~ps (upper) and $t_{\rm Ppr}=-1$ to 10~ps (lower). Except under high-fluence excitation, the PG response nearly equilibrates within $t_{\rm Ppr}=2$~ps. In contrast to the SC case, the PG response exhibits a spatial distribution even under weak excitation. From these data, $A_{\rm PG}$ is defined as $-\langle \Delta R/R \rangle_{0.1-0.4~\mathrm{ps}}$ and plotted in Fig.~\ref{fig_1D} of the main text.

\section{Fluence-dependent 1D analysis using a saturation model}

We summarize the finite-penetration-depth excitation model for the fluence dependence of $\Delta R/R$. If the photoinduced change in the dielectric function $\Delta\epsilon$ at the probe wavelength is described by a depth-dependent profile $g(z)$, then $\Delta R/R$ can be evaluated as the integral of $g(z)$ weighted by the exponential attenuation of the probe. For coaxial Gaussian beams with finite diameters, this expression is extended by integration over the radial coordinate, where $g(r,z)$ is obtained from an effective model that accounts for the spatial dependence of the excitation density:
\begin{equation}
U(r,z)=\mathcal{F}(1-R_{\mathrm{P}})\alpha_{\mathrm{P}}\exp\!\left[-\alpha_{\mathrm{P}}z-\frac{2r^{2}}{\rho_{\mathrm{P}}^{2}}\right],
\end{equation}
where $r$ is the radial distance from the beam center and $R_{\mathrm{P}}$, $\alpha_{\mathrm{P}}$, and $\rho_{\mathrm{P}}$ denote the pump reflectivity, absorption coefficient, and beam radius, respectively. While a phase-shift term can affect $\Delta R/R$, it is neglected here for simplicity; see Ref.~\cite{naseska2018} for a more accurate treatment.

To account for suppression of the SC (PG) state, leading to nonlinear excitation dependence of $\Delta\epsilon$, we adopt a phenomenological saturation model~\cite{kusar2008,naseska2018}. The local amplitude $\Delta\epsilon(r,z)\propto g(r,z)$ is approximated by a piecewise linear function of the locally absorbed energy density $U(r,z)$, with distinct slopes below and above $U_{\mathrm{th}}$:
\begin{eqnarray}
g(r,z) & = & h(U(r,z)),\nonumber \\
h(u) & = & \begin{cases}
u/U_{\mathrm{th}}, & u<U_{\mathrm{th}},\\
1+a\left(u/U_{\mathrm{th}}-1\right), & u\ge U_{\mathrm{th}},
\end{cases}
\end{eqnarray}
where $a$ denotes the relative slope in the normal state and $U_{\mathrm{th}}=\mathcal{F}_{\mathrm{th}}(1-R_{\mathrm{P}})\alpha_{\mathrm{P}}$.

\begin{figure}[htbp]
\includegraphics[width=\columnwidth]{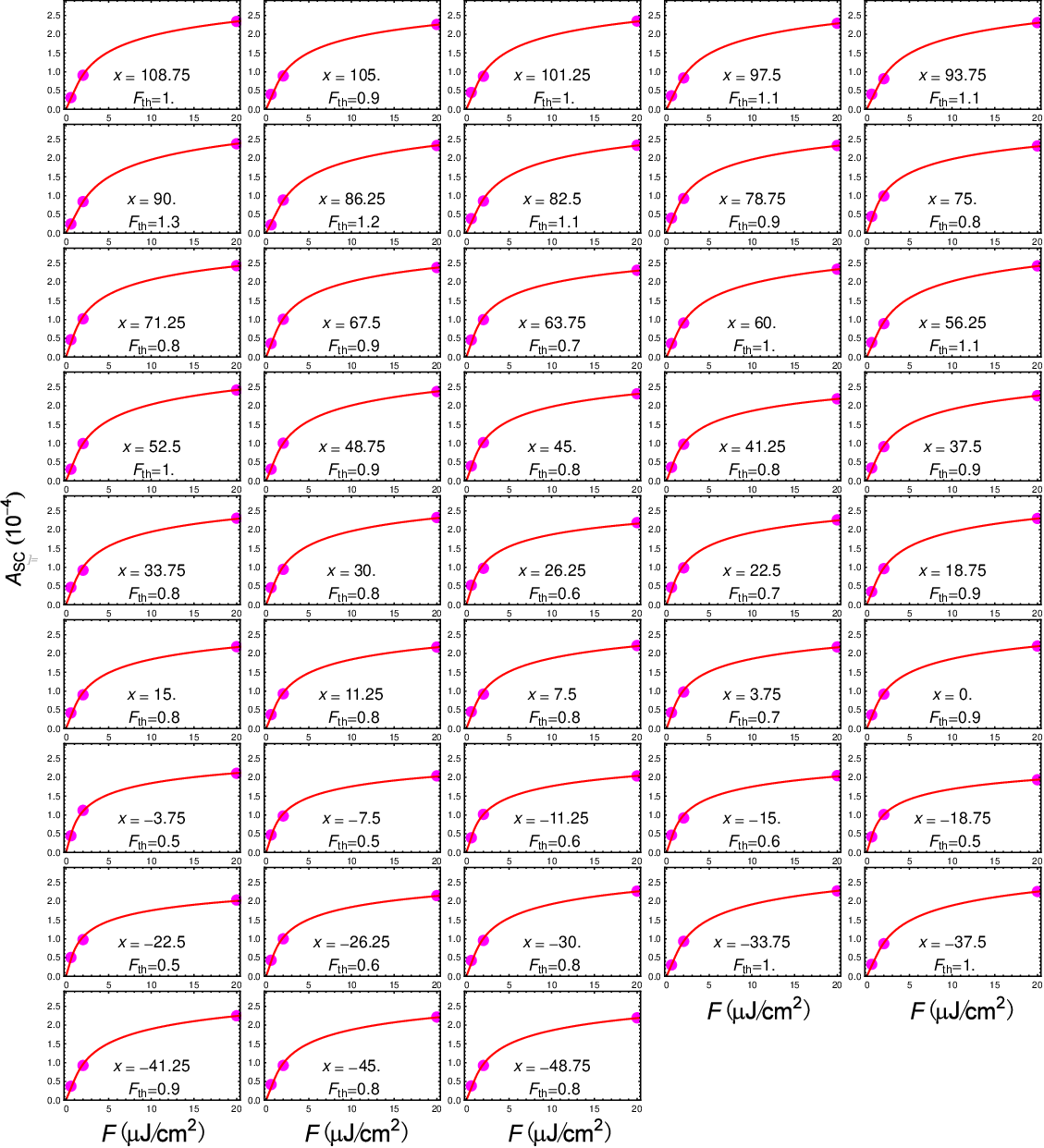}
\caption{(Color online) Fluence dependence of $A_{\rm SC}$ obtained at $T=10~\mathrm{K}$ along the $\sim$160~$\mu$m path indicated in Fig.~\ref{fig_2D}(a). $A_{\rm SC}$ values are extracted from Fig.~\ref{fig_1DSC}.}
\label{fig_FthSC}
\end{figure}

\begin{figure}[htbp]
\includegraphics[width=\columnwidth]{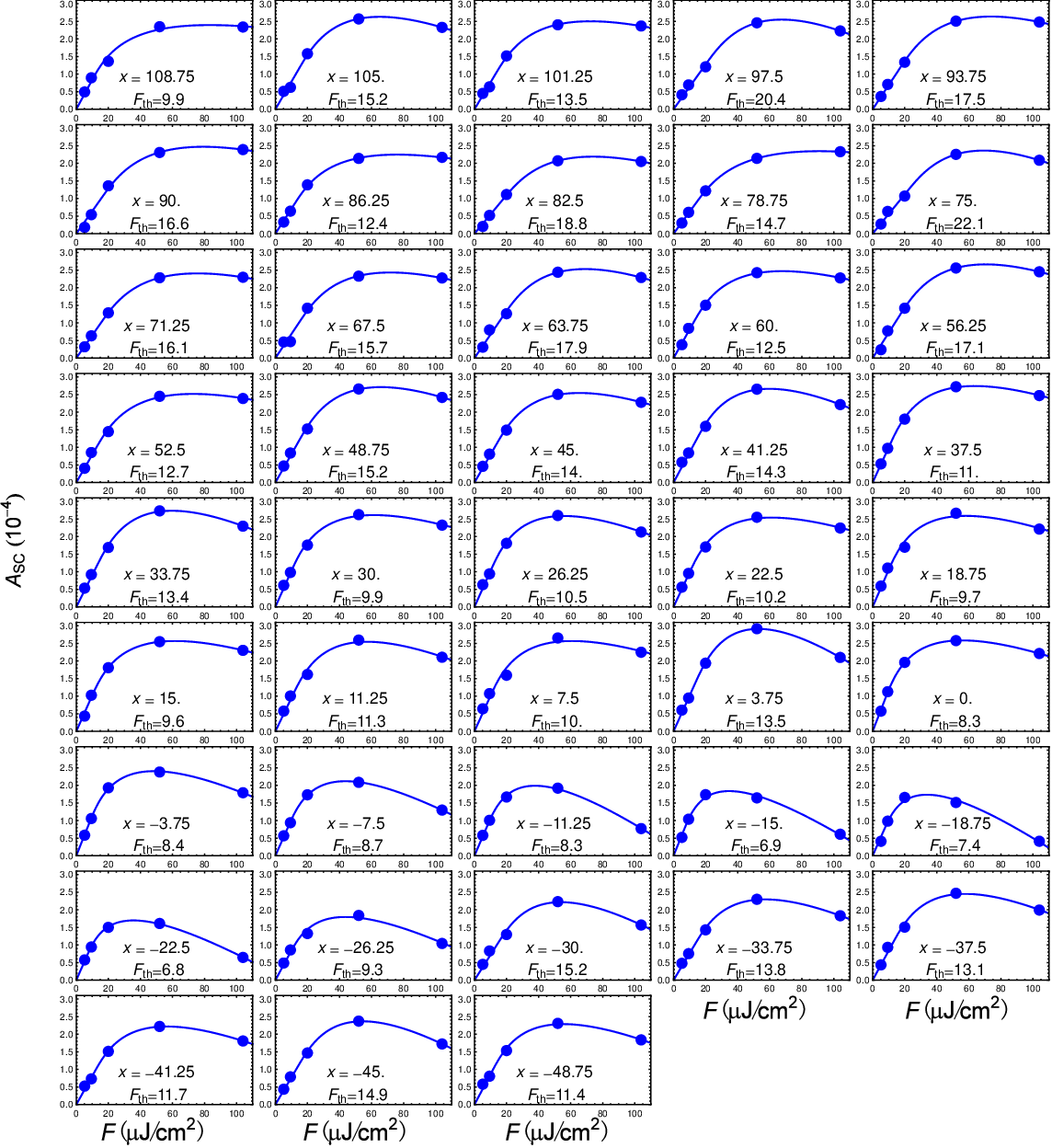}
\caption{(Color online) Fluence dependence of $A_{\rm PG}$ obtained at $T=50~\mathrm{K}$ along the $\sim$160~$\mu$m path indicated in Fig.~\ref{fig_2D}(a). $A_{\rm PG}$ values are extracted from Fig.~\ref{fig_1DPG}.}
\label{fig_FthPG}
\end{figure}

Figures~\ref{fig_FthSC} and \ref{fig_FthPG} present the fluence dependence of $A_{\rm SC}=\langle \Delta R/R \rangle_{2-10~\mathrm{ps}}$ and $A_{\rm PG}=-\langle \Delta R/R \rangle_{0.1-0.4~\mathrm{ps}}$, respectively, obtained at 43 points along the $\sim$160~$\mu$m path in the La-Bi2201 sample. Solid lines show best-fit curves based on the saturation model; the extracted $\mathcal{F}_{\rm th}^{\rm SC,PG}$ correspond to the results in Figs.~\ref{fig_Fth1D}(a),(b) of the main text.

\section{Temperature dependence of $\Delta R/R$ at two characteristic positions}

\begin{figure}[htbp]
\includegraphics[width=\columnwidth]{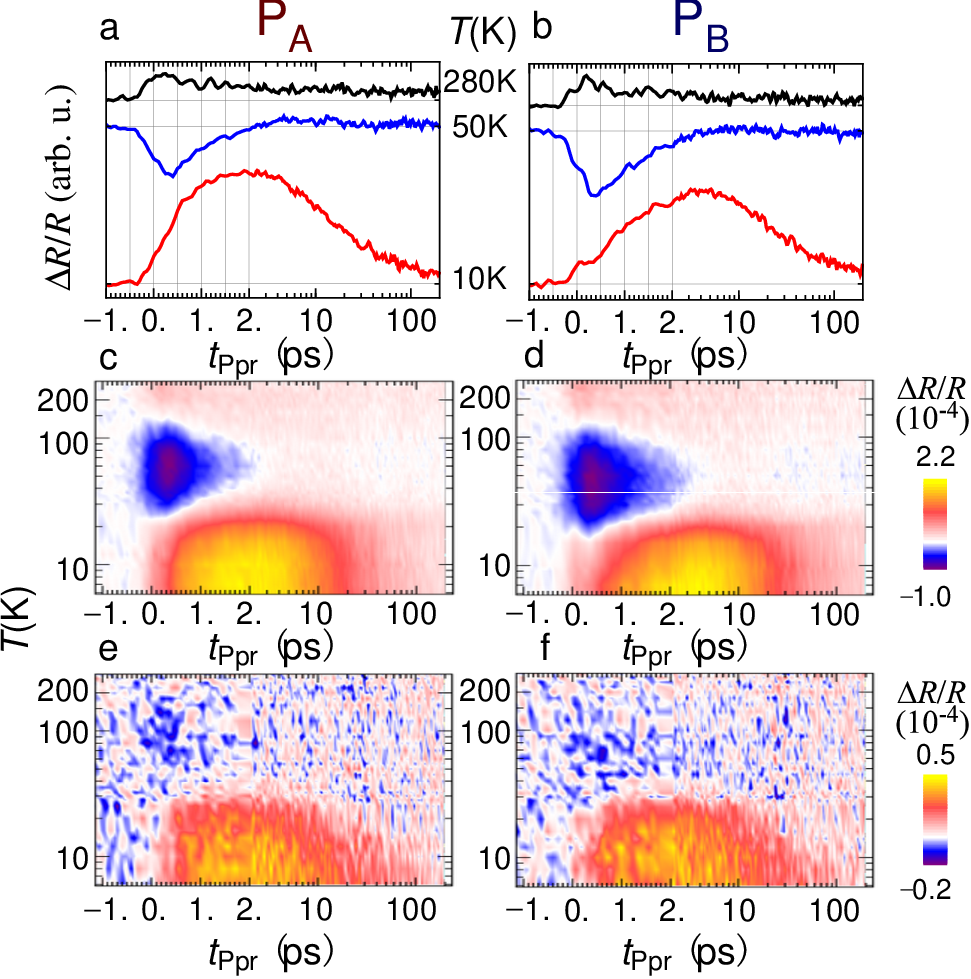}
\caption{(Color online)
Temperature dependence of transient reflectivity at (left) P$_{\rm A}$ and (right) P$_{\rm B}$. (a),(b) $\Delta R/R$ at selected $T$ with ${\mathcal{F}}=15~\mu$J/cm$^{2}$ (shown together in Fig.~\ref{fig_2D}(b) of the main text). Traces are vertically offset for clarity.
(c),(d) Color density plots of $\Delta R/R$ versus time and temperature at ${\mathcal{F}}=15~\mu$J/cm$^{2}$.
(e),(f) Same as (c),(d) but at reduced fluence ${\mathcal{F}}=1.1~\mu$J/cm$^{2}$.
}
\label{fig_TAB}
\end{figure}

In Fig.~\ref{fig_2D}(b) and Fig.~\ref{fig_AB} of the main text, we compare representative $\Delta R/R$ responses at two positions, P$_{\rm A}$ and P$_{\rm B}$ in the La-Bi2201 sample, marked in Fig.~\ref{fig_2D}(a). Figure~\ref{fig_TAB} shows the temperature dependence of $\Delta R/R$ at these locations.

Below $T_{\mathrm{c}}$, $\Delta R/R$ originates from quasiparticle dynamics associated with superconductivity ($\Delta R_{\mathrm{SC}}/R$). A component with the opposite sign, attributed to the pseudogap ($\Delta R_{\mathrm{PG}}/R$), persists up to $\sim$200~K. Near room temperature, $\Delta R/R$ reflects electron-lattice relaxation in the metallic state ($\Delta R_{\mathrm{ER}}/R$). The identification of these components is well established in Bi-based cuprates~\cite{liu2008,giannetti2011,toda2011,coslovich2013,toda2014,akiba2024}.

Figures~\ref{fig_TAB}(a)-\ref{fig_TAB}(d) present results under strong photoexcitation in the superconducting state ($\mathcal{F}_{\rm th}^{\rm SC}<{\mathcal{F}}=15~\mu$J/cm$^2<\mathcal{F}_{\rm th}^{\rm PG}$). Warm-color regions dominating at low $T$ correspond to $\Delta R_{\mathrm{SC}}/R$, while cold-color regions correspond to $\Delta R_{\mathrm{PG}}/R$. The range $T_{\rm c}<T<40~\mathrm{K}$ reflects superconducting fluctuations. Values of $A_{\rm PG}$ in Fig.~\ref{fig_AB}(d) of the main text are derived from $T>40~\mathrm{K}$, excluding the fluctuation region. In contrast, Figs.~\ref{fig_TAB}(e),(f) show results under weak excitation ($\mathcal{F}\approx1.1~\mu$J/cm$^2$), where only the SC QP response is observed; $A_{\rm SC}$ in Fig.~\ref{fig_AB}(c) is derived from $T<45~\mathrm{K}$.

A comparison between Figs.~\ref{fig_TAB}(a) and \ref{fig_TAB}(b) shows that the magnitude of $\Delta R_{\mathrm{PG}}/R$ at P$_{\rm B}$ exceeds that at P$_{\rm A}$, which leads to a slower initial rise of $\Delta R/R$ below $T_{\mathrm{c}}$. This is also evident in the color plots: in Fig.~\ref{fig_TAB}(d), the region associated with $\Delta R_{\mathrm{PG}}/R$ extends to lower temperatures compared with Fig.~\ref{fig_TAB}(c).

\section{Fluence dependence of $\Delta R/R$ at two characteristic positions}
\begin{figure}[htbp]
\includegraphics[width=\columnwidth]{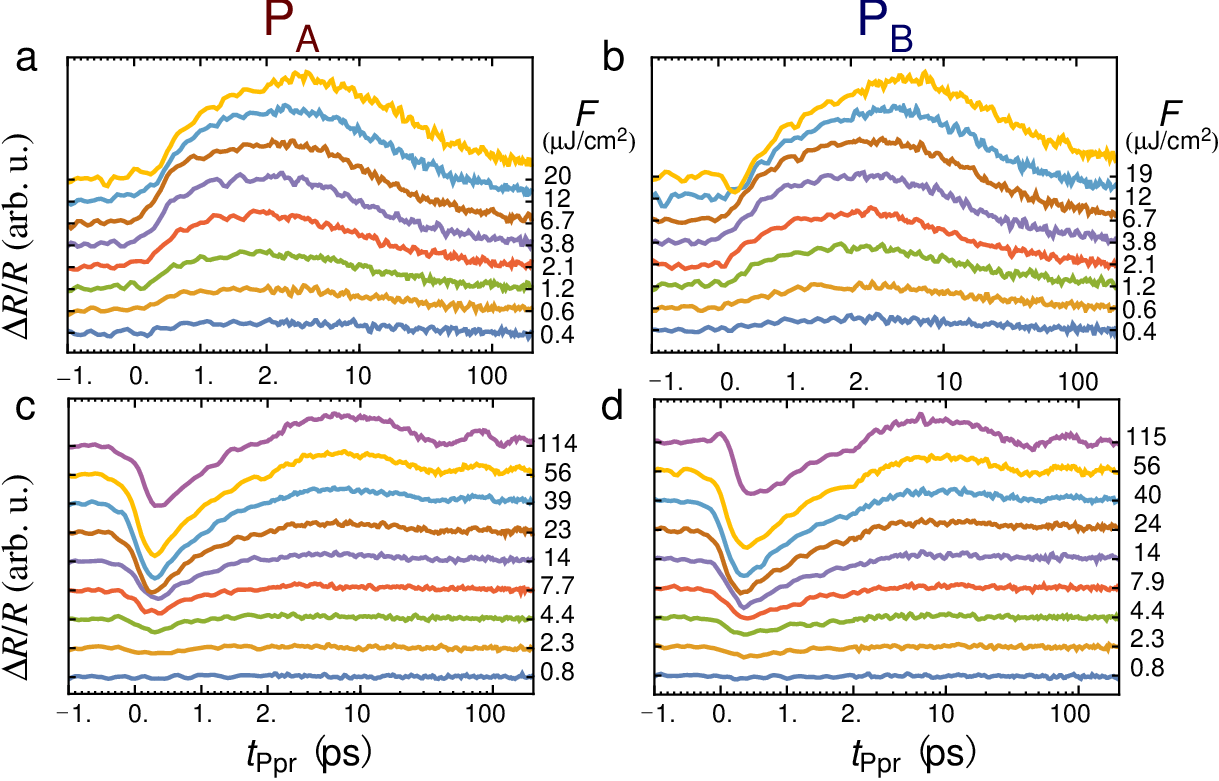}
\caption{(Color online)
Fluence dependence of transient reflectivity at (left) P$_{\rm A}$ and (right) P$_{\rm B}$. (a),(b) $\Delta R/R$ at selected ${\mathcal{F}}$ at $T=10~\mathrm{K}$; (c),(d) at $T=40~\mathrm{K}$. Traces are vertically offset for clarity.
}
\label{fig_FAB}
\end{figure}

Figure~\ref{fig_FAB} presents the fluence dependence of $\Delta R/R$ at P$_{\rm A}$ and P$_{\rm B}$. The upper panels [Figs.~\ref{fig_FAB}(a),(b)] show data at $T=10~\mathrm{K}$, while the lower panels [Figs.~\ref{fig_FAB}(c),(d)] show data at $T=40~\mathrm{K}$. In all cases, clear saturation is observed, indicative of photoinduced phase destruction of the SC ($\Delta R_{\mathrm{SC}}/R$) and PG ($\Delta R_{\mathrm{PG}}/R$) components. The peak shift of $\Delta R/R$ with increasing fluence in Figs.~\ref{fig_FAB}(a),(b) arises from the enhancement of $\Delta R_{\mathrm{PG}}/R$, while $\Delta R_{\mathrm{SC}}/R$ remains nearly constant in the saturated regime. Since $\Delta R_{\mathrm{PG}}/R$ exhibits a higher threshold fluence than $\Delta R_{\mathrm{SC}}/R$, the opposite-sign $\Delta R_{\mathrm{PG}}/R$ becomes dominant above the SC threshold fluence, causing the peak shift.

In this study, the threshold fluences of the SC and PG components, ${\mathcal{F}}{\rm th}^{\rm SC}$ and ${\mathcal{F}}{\rm th}^{\rm PG}$, are defined as the excitation fluence at which the response deviates from the linear behavior observed in the low-fluence regime. As shown in Figs.~\ref{fig_FAB}(a) and (b), the transient response under high-fluence excitation below $T_{\mathrm{c}}$ may include a contribution from the PG component; therefore, the robustness of the evaluation of ${\mathcal{F}}_{\rm th}^{\rm SC}$ needs to be examined.
\begin{figure}[htbp]
\centering
\includegraphics[width=0.8\columnwidth]{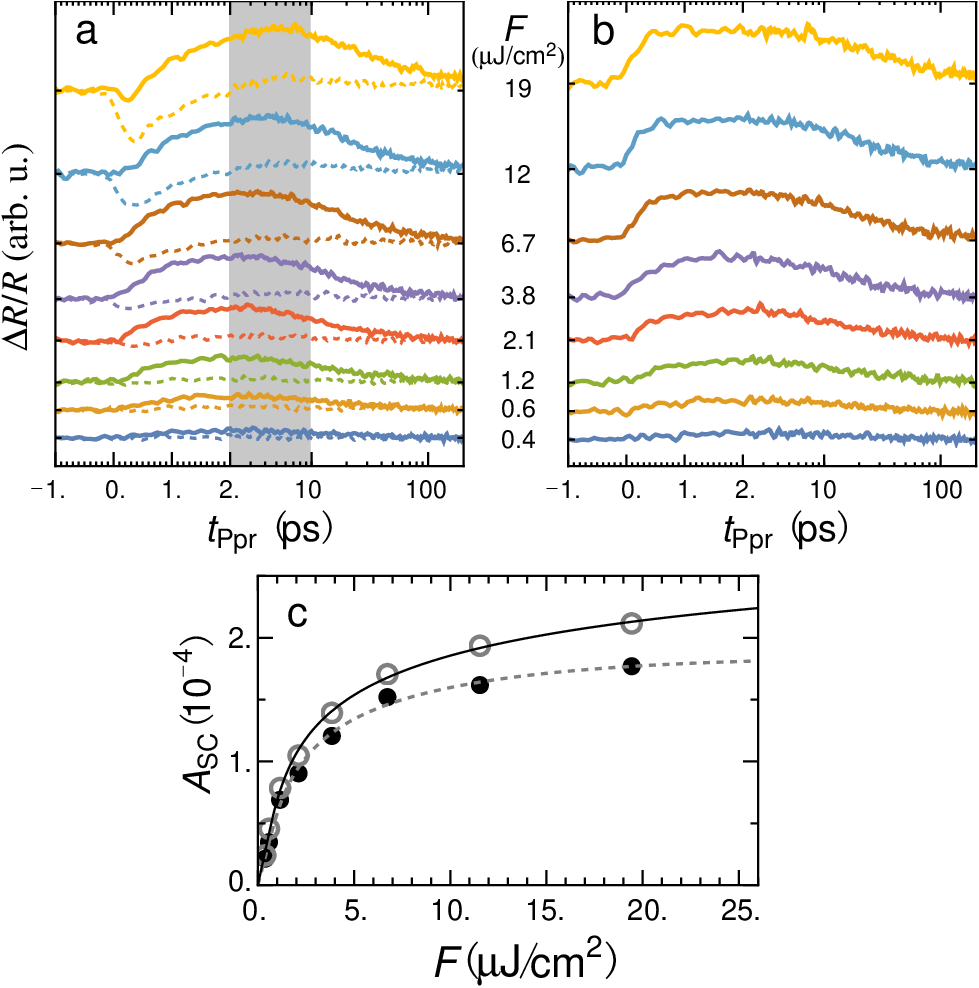}
\caption{(Color online)
(a) Fluence dependence of $\Delta R/R$ at P$_{\rm B}$ measured at 10~K (solid lines) and 30~K (dashed lines). The gray shaded area indicates the time window used to evaluate the SC amplitude, $A_{\rm SC}\equiv\langle \Delta R/R \rangle_{2\text{--}10~\mathrm{ps}}$. 
(b) Fluence dependence of the difference signal, $\Delta R/R$(10~K)$-\Delta R/R$(30~K). 
Traces in (a) and (b) are vertically offset for clarity. 
(c) Fluence dependence of $A_{\rm SC}$ evaluated from the raw data (open circles, $\Delta R/R$(10~K) in (a)) and the subtraction data (filled circles, $\Delta R/R$(10~K)$-\Delta R/R$(30~K) in (b)). 
Solid and dashed lines represent fits based on the finite-penetration-depth excitation model~\cite{kusar2008,naseska2018}.
}
\label{fig_FBsub}
\end{figure}

First, Fig.~\ref{fig_FBsub}(a) shows the fluence dependence of $\Delta R/R$ measured at $T=10$~K (solid lines) and at $T=30$~K (dashed lines), where the PG response is dominant. In the low-fluence regime, the positive $\Delta R/R$ signal at 10~K, characterized by slow relaxation, is significantly larger than the corresponding signal at 30~K. At higher fluences, however, the $\Delta R/R$ signal at 30~K, reflecting the PG response, reaches a comparable peak amplitude. In this regime, a peak shift is observed in $\Delta R/R$ at 10~K, indicating the contribution of the PG component to this signal. On the other hand, in the time window of $t_{\mathrm{Ppr}}>2$~ps, the PG response in $\Delta R/R$ at 30~K has mostly relaxed. Therefore, $A_{\rm SC}\equiv\langle \Delta R/R \rangle_{2\text{--}10~\mathrm{ps}}$ can be regarded as predominantly reflecting the SC component.

Next, Fig.~\ref{fig_FBsub}(b) shows the fluence dependence of the difference signal between $\Delta R/R$ at 10~K and 30~K. As shown in Fig.~\ref{fig_AB}(d), the amplitude of the PG response is nearly temperature-independent in the low-temperature regime, as supported by both the experimental data and the temperature-independent model function. Therefore, the difference signal can be regarded as a transient response in which the SC component is emphasized.

Figure~\ref{fig_FBsub}(c) compares the $A_{\rm SC}$ values obtained from the difference data (filled circles) and those evaluated from the raw data (open circles). Based on the finite-penetration-depth excitation model~\cite{kusar2008,naseska2018}, the threshold fluence of the SC component is estimated to be $0.59 \pm 0.06~\mu\mathrm{J/cm^2}$ from the raw data and $0.65 \pm 0.08~\mu\mathrm{J/cm^2}$ from the difference data. These values agree within the experimental uncertainty.
These results indicate that the admixture of the PG component does not significantly affect the determination of ${\mathcal{F}}_{\rm th}^{\rm SC}$, confirming the robustness of the present threshold analysis.


\end{document}